\documentclass[%
 reprint,
 amsmath,amssymb,
 aps,
 pra,
 citeautoscript,    
]{revtex4-1}

\usepackage{graphicx}
\usepackage{dcolumn}
\usepackage{bm}
\usepackage{braket}
\usepackage[bookmarks,colorlinks,
                      linkcolor=blue,
                      filecolor=blue,
                      citecolor=blue,
                       urlcolor=blue]{hyperref}

\usepackage[colorinlistoftodos,
            textsize=small]{todonotes}

\newcommand{\gap}{\hspace{8mm}}
\newcommand{\reals}{\mathbb{R}}

\newcommand{\idmatrix}{\mathbb{I}}
\newcommand{\vect}[1]{\bm{#1}}
\newcommand{\dV}[1]{d^3\vect{#1}}
\newcommand{\dg}[1]{#1^{\dagger}}
\newcommand{\mean}[1]{\left\langle #1\right\rangle}
\newcommand{\pd}[2]{\frac{\partial #1}{\partial #2}}
\newcommand{\fourierP}[2]{e^{i\vect{#1}\cdot{\vect{#2}}}}
\newcommand{\fourierM}[2]{e^{-i\vect{#1}\cdot{\vect{#2}}}}

\DeclareMathOperator{\Tr}{Tr}
\DeclareMathOperator{\re}{Re}
\DeclareMathOperator{\im}{Im}

\let\epsilon=\varepsilon
\let\phi=\varphi
\let\theta=\vartheta

\begin{document}


\title{All-electron fully relativistic Kohn--Sham theory for solids\\ based on the Dirac--Coulomb Hamiltonian and Gaussian-type functions}

\author{Marius Kadek}
 \email{marius.kadek@uit.no}
\author{Michal Repisky}
\author{Kenneth Ruud}
\affiliation{
 Hylleraas Centre for Quantum Molecular Sciences,\\
 Department of Chemistry, UiT The Arctic University of Norway, Troms\o, Norway
}


\date{\today}

\begin{abstract}
  We present the first full-potential method that solves the fully relativistic
  four-component Dirac--Kohn--Sham equation for materials in the solid state
  within the framework of atom-centered Gaussian-type orbitals (GTOs). Our
  GTO-based method treats one-, two-, and three-dimensional periodic systems on
  an equal footing, and allows for a seamless transition to the methodology
  commonly used in studies of molecules with heavy elements. The scalar
  relativistic effects as well as the spin--orbit interaction are handled
  variationally. The full description of the electron--nuclear potential in the
  core region of heavy nuclei is straightforward due to the local nature of the
  GTOs and does not pose any computational difficulties.  We show how the
  time-reversal symmetry and a quaternion algebra-based formalism can be
  exploited to significantly reduce the increased methodological complexity and
  computational cost associated with multiple wave-function components coupled
  by the spin--orbit interaction. We provide a detailed description of how to
  employ the matrix form of the multipole expansion and an iterative
  renormalization procedure to evaluate the conditionally convergent infinite
  lattice sums arising in studies of periodic systems.  Next, we investigate
  the problem of inverse variational collapse that arises if the Dirac operator
  containing a repulsive periodic potential is expressed in a basis that
  includes diffuse functions, and suggest a possible solution.  Finally, we
  demonstrate the validity of the method on three-dimensional silver halide
  (Ag$X$) crystals with large relativistic effects, and two-dimensional honeycomb
  structures (silicene and germanene) exhibiting the spin--orbit-driven quantum
  spin Hall effect. Our results are well-converged with respect to the basis
  set limit using standard bases developed for molecular calculations, and
  indicate that the common rule of removing basis functions with small
  exponents should not be applied when transferring the molecular basis to
  solids.
\end{abstract}

\maketitle


\section{\label{sec:Introduction}Introduction}

Relativistic effects on band structures and properties of solids containing
heavy elements have for a long time been known to have a significant impact on
both core and valence electrons~\cite{Christensen-PRB-4-3321-1971}. The effects
of relativity on the spectroscopic properties of electrons close to the nuclei
(x-ray spectroscopy) were studied as early as in
1933~\cite{Pincherle-INC-10-344-1933}. In contrast, the importance of
relativistic effects on valence states located close to the Fermi level was not
apparent until 1957~\cite{Mayers-PRSA-241-92-1957} when Mayers observed a large
relativistic contraction of the 6$s$ orbital and a corresponding expansion of the
5$d$ orbitals in heavy elements such as mercury.  Such changes in the size of the
atomic orbitals due to relativity  can lead to dramatic changes in the
structural and physical properties of
solids~\cite{Christensen-PRB-34-5977-1986,Sohnel-AngewChemIntEd-40-4381-2001,Pyykko-CR-88-563-1988}.
For instance, Christensen, Satpathy, and Pawlowska demonstrated that these
relativistic effects are responsible for the stable phase of lead being the
face-centered cubic (fcc) crystal structure, in contrast to the diamond-like
structure adopted by other group 14 elements (C, Si, Ge and
Sn)~\cite{Christensen-PRB-34-5977-1986}. It has also been shown that
relativistic effects need to be included in theoretical models of solids in
order to explains why the ground state of CsAu is insulating and not
metallic~\cite{Christensen-SSC-46-727-1983}. Relativity has also been shown to
significantly increase the voltage of the lead-acid-battery reaction used in
car batteries by 1.7-1.8 V out of the total 2.13
V~\cite{Ahuja-PRL-106-018301-2011}, and lead to a decrease in the melting
temperature of mercury by 105 K~\cite{Calvo-AngewChemIntEd-52-7583-2013},
making mercury the only metal that is liquid at room temperature.

A protruding manifestation of relativity in quantum mechanics -- the
spin--orbit coupling (SOC) -- leads to a splitting of bands in materials
lacking space inversion
symmetry~\cite{Dresselhaus-PR-100-580-1955,Broido-PRB-31-888-1985,Rashba-PLA-129-175-1988}.
These splittings can be remarkably large in transition-metal
dichalcogenides~\cite{Zhu-PRB-84-153402-2011,Heine-ACR-48-65-2014,Rasmussen-JPCC-119-13169-2015,Xu-NatPhys-10-343-2014},
and are then often referred to as ``giant SOC''. SOC plays a paramount role in
the field of
spintronics~\cite{Rashba-PRB-62-R16267-2000,Wolf-Science-294-2001,Zutic-RMP-76-323-2004},
topological insulators~\cite{Kane-PRL-95-226801-2005,Hasan-RMP-82-3045-2010},
and related spin-Hall
effects~\cite{Sinova-PRL-92-126603-2004,Sinova-RMP-87-1213-2015,Kou-JPCL-8-1905-2017}.
SOC has also been shown to open the band gap in two-dimensional honeycomb
systems~\cite{Gmitra-PRB-80-235431-2009,Liu-PRL-107-076802-2011,Xu-PRL-111-136804-2013,Han-NatureNano-9-794-2014}
and change the stable phase of flerovium (Fl, element 114) from fcc to a
hexagonal close packed (hcp) structure~\cite{Hermann-PRB-82-155116-2010}.

Materials exhibiting some of the unique properties mentioned above are
rare~\cite{Bradlyn-Nature-547-298-2017}, however, and the search for novel
materials must be aided by first-principles
calculations~\cite{Bansil-RMP-88-021004-2016}. Modeling spin--orbit-coupled
solid-state systems is far from straightforward, and Kohn--Sham (KS) density
functional theory
(DFT)~\cite{Hohenberg-PR-136-B864-1964,Kohn-PR-140-A1133-1965} is today the
only affordable first-principles method at the fully relativistic level of
theory with variationally included SOC. For such studies, DFT offers a very
favorable compromise between accuracy and computational feasibility.  However,
we note the promising recent works of Sakuma \emph{et
al.}~\cite{Sakuma-PRB-84-085144-2011} and Scherpelz \emph{et
al.}~\cite{Scherpelz-JCTC-12-3523-2016} at the GW level of theory.

A critical choice in the modeling of solids is the representation of the
one-particle basis functions.  There are two major families of basis sets:
local functions (\emph{e.g.} atom-centered orbitals) and plane waves. Plane
waves are ill-suited to capture rapid oscillations of wave functions in regions
close to the nuclei, and are for this reason often combined with
pseudopotentials~\cite{Vanderbilt-PRB-41-7892-1990}. For heavier elements,
these pseudopotentials can be constructed from relativistic all-electron
calculations~\cite{Troullier-PRB-43-1993-1991,Hangele-JCP-136-214105-2012}.
The use of pseudopotentials sacrifices the possibility to model the nodal
structure of the wave functions close to the nuclei and introduces
uncontrollable transferability errors. This makes all-electron methods in some
cases the preferred method, \emph{e.g.} for calculations of nuclear magnetic
resonance (NMR) shifts~\cite{Kim-JACS-132-16825-2010}.

Relativistic all-electron calculations are possible using the relativistic
Korringa--Kohn--Rostoker (KKR)
method~\cite{Takada-PTP-36-224-1966,Strange-JPhysC-17-3355-1984,Wang-PRB-46-9352-1992,Der-PRB-54-11187-1996}
or by extending Slater's augmented plane-wave (APW)
method~\cite{Slater-PR-51-846-1937} to the Dirac
Hamiltonian~\cite{Loucks-PR-139-A1333-1965,Mattheiss-PR-151-450-1966}. The APW
method divides space into spheres centered at atoms and an interstitial region,
and requires solving a secular energy-dependent equation for each band to match
KS orbitals at boundaries of the spheres. This approach results in equations
with a nonlinear dependence on energies. The method is very accurate, but
computationally expensive. To mitigate the computational cost, the APW method
can be linearized~\cite{Andersen-PRB-12-3060-1975,Sjostedt-SSC-114-15-2000},
leading to the linear-APW (LAPW) and linear muffin-tin orbitals (LMTO) methods,
enabling the use of a full potential for all electrons. The LMTO approach has
been extended to the relativistic
domain~\cite{Christensen-JPF-8-L51-1978,Godreche-JMMM-29-262-1982,Ebert-PRB-38-9390-1988,Ebert-JAP-63-3052-1988}.
A relativistic extension of LAPW was first developed by MacDonald, Picket and
Koelling~\cite{MacDonald-JPhC-13-2675-1980} and later by Wimmer \emph{et
al.}~\cite{Wimmer-PRB-24-864-1981}. MacDonald \emph{et al.} included SOC by a
two-step variational method, the so-called second-variational approach,
\emph{i.e.} as a post processing to the spin-non-polarized scalar-relativistic
self-consistent procedure. This process is performed on a smaller set of
scalar-relativistic eigenfunctions, thus reducing the computational effort
considerably. The second-variational approach was later extended and
implemented in some of the modern program
packages~\cite{Schwarz-CPC-147-71-2002,wien2k,Kurz-PRB-69-024415-2004}, where
the second-variational inclusion of SOC can be employed both self-consistently
as well as non-self-consistently. 

Both the full-potential LMTO and LAPW methods suffer from limitations when
treating systems with deep-lying valence and extended core
states~\cite{Blaha-PRB-46-1321-1992}. If SOC is included, severe convergence
problems can be encountered~\cite{Nordstrom-PRB-63-035103-2000}. These
limitations are due to the insufficient flexibility of the finite
scalar-relativistic basis set for describing Dirac $p_{1/2}$ states in the core
region~\cite{MacDonald-JPhC-13-2675-1980,Nordstrom-PRB-63-035103-2000}.
Convergence is achieved when the basis is augmented by Dirac $p_{1/2}$ local
orbitals in the second variational
step~\cite{Kunes-PRB-64-153102-2001,Carrier-PRB-70-035212-2004,Huhn-PRM-1-033803-2017}.
Huhn and Blum carried out a benchmark study and a comparison of various LAPW
strategies for the evaluation of the SOC
contribution~\cite{Huhn-PRM-1-033803-2017}.

More recently, the linearized methods were generalized by Bl{\"o}chl to include
the pseudopotential approximation, establishing the projector augmented wave
(PAW) technique~\cite{Blochl-PRB-50-17953-1994,Kresse-PRB-59-1758-1999}. PAW
introduces pseudopotentials as a well-defined approximation, and hence brings
transferability errors under control, enabling all-electron  calculations of
properties in a pseudopotential framework. However, the complexity of the PAW
approach makes its extension to {\it e.g.\/} include hybrid density functionals
and the study of response properties difficult. A fully relativistic PAW method
for both Dirac-type (four-component) and Pauli-type (two-component) equations was
formulated by Dal Corso~\cite{DalCorso-PRB-82-075116-2010}. 

An alternate strategy to the use of  plane waves, is to expand the KS orbitals
in a set of local functions. Such full-potential methods employing numerical
orbitals have been extended to include scalar relativistic
corrections~\cite{Suzuki-JPSJ-69-532-2000,Blum-CPC-180-2175-2009}, as well as
four-component (4c)
SOC~\cite{Suzuki-JPSJ-68-1982-1999,Koepernik-PRB-59-1743-1999,Opahle-phdthesis-2001,chapter-Eschrig-2004}.
Alternatively, basis functions can be constructed by placing analytic
Slater-type orbitals (STOs) or Gaussian-type orbitals (GTOs) on atomic centers.
two-component (2c) techniques using STOs were implemented by Philipsen \emph{et
al.}~\cite{Philipsen-PRB-56-13556-1997,Philipsen-PRB-61-1773-2000} and Zhao
\emph{et al.}~\cite{Zhao-JCP-144-044105-2016} Relativistic calculations on
solids with GTOs were reported with scalar-relativistic
corrections~\cite{Geipel-CPL-273-62-1997,Peralta-JCP-122-084108-2005}, as well
as approximate 2c schemes solving Pauli-type
equations~\cite{Wang-PRB-9-4897-1974,Mainkar-PRB-53-3692-1996}, or approaches
based on the Douglass--Kroll--Hess
Hamiltonian~\cite{Jones-PRB-61-4644-2000,Boettger-PRB-62-7809-2000}.  While
calculations that include scalar-relativistic corrections on solids are
common~\cite{Suzuki-JPSJ-69-532-2000,Blum-CPC-180-2175-2009,Geipel-CPL-273-62-1997,Peralta-JCP-122-084108-2005},
extending nonrelativistic implementations by SOC poses severe methodological
challenges due to the appearance of multicomponent spinor structure of the wave
functions as well as the need to use complex algebra.

Here, we present the first fully relativistic all-electron full-potential
GTO-based method directly solving the 4c Dirac--Kohn--Sham (DKS) equation for
periodic systems while treating both the scalar relativistic effects and SOC
variationally during the self-consistent optimization procedure. Thus, the
approach enables studies of relativistic effects in solids containing elements
from the entire periodic table in a consistent manner without the use of the
pseudopotential approximation. The variational treatment of SOC is mandatory in
studies of materials containing heavy elements, where SOC splittings are of the
same magnitude as the effects of the crystal potential, and for which the
evaluation of perturbational or non-self-consistent SOC can be
insufficient~\cite{Soven-PR-137-A1706-1965,Sakuma-PRB-84-085144-2011,Huhn-PRM-1-033803-2017}.
We will demonstrate that GTOs are a convenient and computationally efficient
approach for full-potential relativistic calculations. The local nature of the
GTOs makes them amendable to highly efficient linear scaling techniques, as
GTOs better reflect the decay properties of operators and density
matrices~\cite{Goedecker-RMP-71-1085-1999}. In addition, because periodicity is
embedded explicitly in the local basis, systems that are periodic in one or two
dimensions (polymers and slabs) can be studied using atom-centered GTOs while
avoiding the requirement to repeat the polymer or slab in the non-periodic
dimensions~\cite{Heine-ChemSocRev-43-6537-2014}. This eliminates the concern in
calculations on such systems using plane waves that there will be spurious self
interactions between the system studied and its periodic images.  In contrast
to LMTO and LAPW, GTOs can treat both core and valence electrons on an equal
footing, the quality being independent of a fixed linearization energy.
Furthermore, here we will demonstrate that a systematic convergence to the
basis limit is possible with GTOs, and discuss specific problems arising in the
nonrelativistic and 4c cases if the diffuse functions are included in a local
basis.

The presented 4c method builds on a transparent and efficient quaternion
algebra-based formulation of time-reversal-symmetric operators in real and
reciprocal space employing a Kramers-restricted kinetically balanced basis. We
implemented this method into the 4c \textsc{ReSpect} program
package~\cite{respect5}, and utilized integral screening techniques based on
quaternion algebra~\cite{Konecny-JCP-149-204104-2018,Repisky-integralPaper}.
The implementation exploits the full $\vect{k}$-point sampling of the first
Brillouin zone, and allows to use irreducible unit cells for all lattice
structures. Our approach builds on previous nonrelativistic methodologies for
handling periodic systems with GTOs. This includes the pioneering works of
Pisani, Dovesi and
coworkers~\cite{Pisani-IJQC-17-501-1980,Dovesi-PRB-28-5781-1983}, and the more
recent implementations of Towler, Zupan and
Caus\'{a}~\cite{Towler-CPC-98-181-1996} and of {\L}azarski, Burow and
Sierka~\cite{Lazarski-JCTC-11-3029-2015}.

The rest of the paper is organized as follows. In Sec.~\ref{sec:Theory}, we
establish the main principles of our 4c GTO-based method for the solid state.
In Sec.~\ref{sec:GeneralFramework}, we concentrate on the general formulation
of the working equations, in Sec.~\ref{sec:DensityAndDensityMatrix}, we define
the 4c density and the density matrix in real-space GTOs,
Sec.~\ref{sec:TimeReversalSymmetry} shows consequences of the time-reversal
symmetry on the structure of operators in both real space and reciprocal space,
and these concepts are further developed in Sec.~\ref{sec:QuaternionOperators}
in a quaternion formulation.  In Sec.~\ref{sec:CoulombPotentialAndEnergy}, we
derive how the Coulomb potential and energy are evaluated using the 4c
real-space GTOs, before we in
Sec.~\ref{sec:TreatmentOfElectrostaticLatticeSums} analyze the problem of the
long-range electrostatic lattice sums, and describe its solutions within our
theoretical framework. In Sec.~\ref{sec:ExchangeCorrelationContribution}, we
derive the exchange--correlation contributions. Practical implementation
details and approximations required in realistic calculations are described in
Sec.~\ref{sec:ImplementationDetails}. In Sec.~\ref{sec:NonrelativisticTheory},
we discuss problems emerging in the nonrelativistic solid-state calculations
associated with diffuse functions in local basis sets, and in
Sec.~\ref{sec:RelativisticTheory}, we outline a new complication that arises in
our relativistic method if we include  diffuse functions in the basis. Results
for the silver halide crystals and 2D hexagonal structures are shown and
discussed in Sec.~\ref{sec:Results}, before we in Sec.~\ref{sec:Conclusion}
give some concluding remarks and an outlook.

\section{\label{sec:Theory}Theory}

\subsection{\label{sec:GeneralFramework}General framework}

In this section, we outline the basic GTO-based scheme we use to solve the 4c
DKS equations for periodic systems. Unless otherwise stated, we employ atomic
units, setting the elementary charge $e$, the electron rest mass $m_e$ and
reduced Planck's constant $\hbar$ to unity. Throughout this paper, Einsteins's
implicit summation over repeated indices is assumed.

The fundamental building units of the presented theory are the scalar
atom-centered normalized primitive Cartesian
GTOs~\cite{Boys-PRSA-200-542-1950,Obara-JCP-84-3963-1986}
\begin{equation}
	g_{\mu}(\vect{r}) \equiv \mathcal{N}(x-A_x)^{l_x}(y-A_y)^{l_y}(z-A_z)^{l_z} e^{-\alpha(\vect{r}-\vect{A})^2},
\label{eq:cartesianGTOs}
\end{equation}
where $\mathcal{N}$ is the normalization constant, $\alpha$ is the
Gaussian exponent, $\vect{l}\equiv(l_x,l_y,l_z)$ are the Cartesian angular
momenta, and $\vect{A}$ and $\vect{r}$ are the nuclear and electron
coordinates, respectively.  Basis representations of the solutions to the DKS
equations are constructed in three steps. First, 4c basis bispinors
$\chi_{\mu}$ for a reference unit cell are formed
\begin{equation}
	\chi_{\mu}(\vect{r}) \equiv \begin{pmatrix}
		\chi^L_{\mu}(\vect{r}) & 0_2 \\
		0_2 & \chi^S_{\mu}(\vect{r})
	\end{pmatrix},
\label{eq:AObasis4c}
\end{equation}
using 2c spinors $\chi^L_{\mu}$ and $\chi^S_{\mu}$ defined for the so-called
large ($L$) and small ($S$) components, respectively, as
\begin{subequations}
\label{eq:AObasis4cGroup}
\begin{align}
	\chi^L_{\mu}(\vect{r}) &\equiv \idmatrix_2 \otimes g_{\mu}(\vect{r}), \label{eq:AObasis4cLL}\\
	\chi^S_{\mu}(\vect{r}) &\equiv \frac{1}{2c}(\vect{\sigma}\cdot\vect{p}) g_{\mu}(\vect{r}), \label{eq:AObasis4cSS}
\end{align}
\end{subequations}
where $\idmatrix_2$ is the $2\times 2$ identity matrix, $\vect{\sigma}$ are the
Pauli matrices, $\vect{p}\equiv-i\vect{\nabla}$ is the electron momentum
operator, and $c$ is the speed of light. The construction of the
small-component basis in Eq.~\eqref{eq:AObasis4cSS} utilizes the restricted
kinetically balanced (RKB) condition which is essential to achieve
variationally stable 4c solutions in a finite
basis~\cite{Stanton-JCP-81-1910-1984}.  Second, the basis for periodic systems
is obtained by translating $\chi_{\mu}$ from the reference unit cell to the
unit cell $\vect{m}$ as
\begin{equation}
	\chi_{\mu\vect{m}}(\vect{r}) \equiv \chi_{\mu}(\vect{r}-\vect{m}),
\label{eq:AObasisPeriodic}
\end{equation}
where the unit cell position vector $\vect{m}$ is
\begin{equation}
	\vect{m} = m^i \vect{a}_i, \gap m^i\in\mathbb{Z}, i=1,\ldots,d.
\label{eq:defmvec}
\end{equation}
Here, $\mathbb{Z}$ denotes the field of integers, $d$ is the number of periodic
dimensions, and $\vect{a}_i$ are the primitive vectors that constitute a
Bravais lattice. Since all unit cells are equivalent, we choose the central
unit cell $\vect{m}=\vect{0}$ to be the fixed reference unit cell.  Third,
symmetry-adapted Bloch functions for each $\vect{k}$ point from the first
Brillouin zone $\mathcal{K}$ are constructed from the real-space GTOs as the
Fourier series
\begin{equation}
	\phi_{\mu}(\vect{k};\vect{r}) = \frac{1}{\sqrt{|\mathcal{K}|}} \sum_{\vect{m}} \fourierP{k}{m} \chi_{\mu\vect{m}}(\vect{r}),
\label{eq:kSpaceBasis}
\end{equation}
where the infinite lattice sum is over the whole Bravais lattice.
$|\mathcal{K}|$ is the volume of the primitive reciprocal unit cell (first
Brillouin zone), and enters the normalization constant to ensure an approximate
normalization of the Bloch functions. The symmetry-adapted functions in
Eq.~\eqref{eq:kSpaceBasis} satisfy the Bloch condition
\begin{equation}
	\phi_{\mu}(\vect{k};\vect{r}+\vect{m}) = \fourierP{k}{m}\phi_{\mu}(\vect{k};\vect{r}),
\label{eq:BlochCondition}
\end{equation}
by construction, and $\phi_{\mu}(\vect{k};\vect{r})$ can thus be used as basis
functions that block-diagonalize a translationally invariant Hamiltonian.

Our aim is to solve the 4c DKS equations
\begin{equation}
	\hat{F}\psi_p(\vect{k};\vect{r}) = \epsilon_p(\vect{k})\psi_p(\vect{k};\vect{r}),
\label{eq:DKSinCOs}
\end{equation}
for each band $p$. Here $\epsilon_p(\vect{k})$ and $\psi_p(\vect{k};\vect{r})$
are the energy and the crystalline orbital (CO) of the $p$-th band,
respectively, and $\hat{F}$ is the 4c Fock operator
\begin{equation}
	\hat{F} = \begin{pmatrix}
		V(\vect{r}) & c\vect{\sigma}\cdot\vect{p} \\
		c\vect{\sigma}\cdot\vect{p} & V(\vect{r}) - 2c^2
	\end{pmatrix},
\label{eq:Fock4cOperator}
\end{equation}
consisting of the one-electron Dirac Hamiltonian~\cite{Dirac-PRSA-117-610-1928}
and the potential $V(\vect{r})$, which in the context of KS DFT contains the
mean-field Coulomb potential and the exchange--correlation
potential~\cite{Kutzelnigg-IJQC-25-107-1984,Saue-JCC-23-814-2002,Komorovsky-JCP-128-104101-2008}.
Such an approach approximates the two-electron interaction with an
instantaneous Coulomb operator, neglecting the relativistic corrections to the
electron--electron interaction.  We expand the solutions
$\psi_p(\vect{k};\vect{r})$ of Eq.~\eqref{eq:DKSinCOs} in terms of the Bloch
functions in Eq.~\eqref{eq:kSpaceBasis}:
\begin{equation}
	\psi_p(\vect{k};\vect{r}) = \phi_{\mu}(\vect{k};\vect{r}) c^{\mu}_{\phantom{a}p}\, (\vect{k}),
\label{eq:COexpansion}
\end{equation}
where $c^{\mu}_{\phantom{a}p}(\vect{k})$ are the 4c CO expansion coefficients.
Inserting the expansions in Eqs.~\eqref{eq:COexpansion}
and~\eqref{eq:kSpaceBasis} into Eq.~\eqref{eq:DKSinCOs}, multiplying the
equation with $\dg{\phi}_{\mu'}(\vect{k};\vect{r})$ from the left and
integrating over spatial coordinates $\vect{r}$, yields the matrix form of the
DKS equation in reciprocal space
\begin{equation}
	F(\vect{k})c(\vect{k}) = S(\vect{k})c(\vect{k})\epsilon(\vect{k}),
\label{eq:DKSinMatrixForm}
\end{equation}
where $\epsilon(\vect{k})$ is the diagonal matrix of the band energies.
$F(\vect{k})$ and $S(\vect{k})$ are reciprocal-space forms of the Fock and
overlap matrices, respectively (see Appendix~\ref{app:TranslationSymmetry}):
\begin{subequations}
\label{eq:kSpaceGroupFandS}
\begin{align}
	F_{\mu\mu'}(\vect{k}) &= \sum_{\vect{m}} \fourierP{k}{m} F_{\mu\vect{0},\mu'\vect{m}}, \label{eq:kSpaceFock}\\
	S_{\mu\mu'}(\vect{k}) &= \sum_{\vect{m}} \fourierP{k}{m} S_{\mu\vect{0},\mu'\vect{m}}, \label{eq:kSpaceOverlap}
\end{align}
\end{subequations}
and
\begin{subequations}
\label{eq:rSpaceGroupFandS}
\begin{align}
	F_{\mu\vect{0},\mu'\vect{m}} &= \int_{\reals^3} \dg{\chi}_{\mu\vect{0}}(\vect{r}) \hat{F}\chi_{\mu'\vect{m}}(\vect{r}) \dV{r}, \label{eq:rSpaceFock}\\
	S_{\mu\vect{0},\mu'\vect{m}} &= \int_{\reals^3} \dg{\chi}_{\mu\vect{0}}(\vect{r}) \chi_{\mu'\vect{m}}(\vect{r}) \dV{r}. \label{eq:rSpaceOverlap}
\end{align}
\end{subequations}
We have here exploited the translational invariance of the Fock operator, which
allows us to consider only the reference unit cell $\vect{m}=\vect{0}$ for the
bra function $\dg{\chi}_{\mu\vect{0}}$, and to solve
Eq.~\eqref{eq:DKSinMatrixForm} independently for each $\vect{k}$.  Finally, we
express the real-space integrals in Eqs.~\eqref{eq:rSpaceGroupFandS} utilizing
Eqs.~\eqref{eq:AObasis4c} and~\eqref{eq:AObasis4cGroup} to obtain the 4c matrix
forms for $F_{\mu\vect{0},\mu'\vect{m}}$ and $S_{\mu\vect{0},\mu'\vect{m}}$:
\begin{align}
	F_{\mu\vect{0},\mu'\vect{m}} &= \begin{pmatrix}
		\mathcal{V}^{LL} & \mathcal{T} \\
		\mathcal{T} & \frac{1}{4c^2}\mathcal{V}^{SS}-\mathcal{T}
	\end{pmatrix}_{\mu\vect{0},\mu'\vect{m}}, \label{eq:rSpaceFock4cMat}\\
	S_{\mu\vect{0},\mu'\vect{m}} &= \begin{pmatrix}
		\mathcal{S} & 0_2 \\
		0_2 & \frac{1}{2c^2}\mathcal{T}
	\end{pmatrix}_{\mu\vect{0},\mu'\vect{m}}, \label{eq:rSpaceOverlap4cMat}
\end{align}
where the indices $\mu\vect{0},\mu'\vect{m}$ are applied to each element of the
matrices individually and
\begin{subequations}
\label{eq:rSpace2cGroup}
\begin{align}
	\mathcal{S}_{\mu\vect{0},\mu'\vect{m}} &= \idmatrix_2\otimes\int_{\reals^3} g_{\mu\vect{0}}(\vect{r}) g_{\mu'\vect{m}}(\vect{r}) \dV{r}, \label{eq:rSpace2cSMat}\\
	\mathcal{T}_{\mu\vect{0},\mu'\vect{m}} &= \idmatrix_2\otimes\int_{\reals^3} g_{\mu\vect{0}}(\vect{r}) \frac{p^2}{2} g_{\mu'\vect{m}}(\vect{r}) \dV{r}, \label{eq:rSpace2cTMat}\\
	\mathcal{V}^{LL}_{\mu\vect{0},\mu'\vect{m}} &= \idmatrix_2\otimes\int_{\reals^3} g_{\mu\vect{0}}(\vect{r}) V(\vect{r}) g_{\mu'\vect{m}}(\vect{r}) \dV{r}, \label{eq:rSpace2cVMat}\\
	\mathcal{V}^{SS}_{\mu\vect{0},\mu'\vect{m}} &= \int_{\reals^3} \left[(\vect{\sigma}\cdot\vect{p})g_{\mu\vect{0}}(\vect{r})\right]^{\dagger}V(\vect{r})\left[(\vect{\sigma}\cdot\vect{p})g_{\mu'\vect{m}}(\vect{r})\right] \dV{r}. \label{eq:rSpace2cWMat}
\end{align}
\end{subequations}
Integrals over the GTOs in Eqs.~\eqref{eq:rSpace2cGroup} are evaluated
analytically using the recurrence scheme of Obara and
Saika~\cite{Obara-JCP-84-3963-1986,interest}.  If we now let
\begin{align}
	T &= \begin{pmatrix}
		0_2 & \mathcal{T} \\
		\mathcal{T} & -\mathcal{T}
	\end{pmatrix}, & V &= \begin{pmatrix}
		\mathcal{V}^{LL} & 0_2 \\
		0_2 & \frac{1}{4c^2}\mathcal{V}^{SS}
	\end{pmatrix},
\label{eq:rSpaceTmatAndVmat4comp}
\end{align}
be the 4c kinetic energy matrix and the potential matrix, respectively, where
we have omitted the $\mu\vect{0},\mu'\vect{m}$ indices, the DKS Fock matrix in
Eq.~\eqref{eq:rSpaceFock4cMat} can be partitioned as:
\begin{equation}
	F = T + V = T + J + V^{\text{XC}}.
\label{eq:rSpaceFock4cMatParts}
\end{equation}
Here, $J$ is the Coulomb and $V^{\text{XC}}$ the exchange--correlation
contribution to the potential matrix $V$  (the evaluation of these
contributions will be discussed in more detail in
Secs.~\ref{sec:CoulombPotentialAndEnergy}
and~\ref{sec:ExchangeCorrelationContribution}, respectively). The Coulomb
matrix $J$ contains both the electron-nuclear interaction and the Hartree
mean-field interaction term. The exact exchange matrix required for
Hartree--Fock theory or hybrid DFT is omitted in this work.

Within the framework of DFT, Eq.~\eqref{eq:DKSinMatrixForm} must be solved
self-consistently, since $V$ contains the mean-field potential as well as the
exchange--correlation potential, both of which depend on the electron density
and its gradients and which are constructed from the COs
$\psi_i(\vect{k};\vect{r})$.  Eq.~\eqref{eq:DKSinMatrixForm} is solved in an
iterative manner: its solutions are used to build a new Fock matrix $F$,
Eq.~\eqref{eq:DKSinMatrixForm} is then solved for this updated potential until
convergence is reached.

\subsection{\label{sec:DensityAndDensityMatrix}Density and density matrix}

In this section, we formulate the real-space 4c reduced one-electron density
matrix $D^{\mu\vect{m},\mu'\vect{0}}$ and the electron density $\rho_\text{e}$
for periodic systems that are used in practice for the construction of the Fock
matrix [Eq.~\eqref{eq:rSpaceFock4cMat}] instead of $\psi_i(\vect{k};\vect{r})$.

The reciprocal-space density matrix expressed in terms of COs is a diagonal
matrix, where the diagonal elements form an occupation vector $f_p(\vect{k})$
for each band $p$. $f_p(\vect{k})$ is a zero-temperature limit of the
Fermi--Dirac distribution
\begin{equation}
	f_p(\vect{k}) = \frac{1}{e^{\beta\left(\epsilon_p(\vect{k})-\mu\right)} + 1} \overset{\beta\rightarrow\infty}{\longrightarrow} \theta\left(\mu-\epsilon_p(\vect{k})\right),
\label{eq:FermiDirac}
\end{equation}
where $\mu$ is the Fermi level chemical potential, $\beta$ is the inverse
temperature, and $\theta$ is the Heaviside step function. Bands corresponding
to positronic (negative-energy) states are left vacant (see
Sec.~\ref{sec:RelativisticTheory}).  If we let $f(\vect{k})$ denote the
diagonal matrix of occupation numbers, we can write the $\vect{k}$-space
density matrix in its block-diagonal form as
\begin{align}
	D(\vect{k},\vect{k}') &= \delta(\vect{k}-\vect{k}')D(\vect{k}), \label{eq:kSpaceDmat1}\\
	D(\vect{k}) &= c(\vect{k})f(\vect{k})\dg{c}(\vect{k}), \label{eq:kSpaceDmat2}
\end{align}
where $\delta$ is the Dirac delta function.  Inverting the Fourier series in
Eq.~\eqref{eq:kSpaceBasis}, gives
\begin{equation}
	\chi_{\mu\vect{m}}(\vect{r}) = \frac{1}{\sqrt{|\mathcal{K}|}} \int_{\mathcal{K}} \fourierM{k}{m} \phi_{\mu}(\vect{k};\vect{r}) \dV{k},
\label{eq:kSpaceBasisInverse}
\end{equation}
which we use together with Eq.~\eqref{eq:kSpaceDmat1} to obtain the real-space
density matrix as a quadrature
\begin{equation*}
	D^{\mu\vect{m},\mu'\vect{m}'} = \frac{1}{|\mathcal{K}|} \int_{\mathcal{K}}e^{i\vect{k}\cdot(\vect{m}-\vect{m}')} D^{\mu\mu'}(\vect{k}) \dV{k},
\end{equation*}
where $D^{\mu\mu'}(\vect{k})$ are elements of the matrix defined in
Eq.~\eqref{eq:kSpaceDmat2}. In practice, it is enough to restrict ourselves
only to nonequivalent elements (see Appendix~\ref{app:TranslationSymmetry}):
\begin{equation}
	D^{\mu\vect{m},\mu'\vect{0}} = \frac{1}{|\mathcal{K}|} \int_{\mathcal{K}}\fourierP{k}{m} D^{\mu\mu'}(\vect{k}) \dV{k}.
\label{eq:rSpaceDmatQuadrature}
\end{equation}

The electron charge density can be evaluated as (the minus sign is for the
electron charge)
\begin{equation}
	\rho_{\text{e}}(\vect{r}) \equiv -\sum_p \int_{\mathcal{K}} \Tr\left[ \psi_p(\vect{k};\vect{r})\dg{\psi}_p(\vect{k};\vect{r}) f_p(\vect{k})\right]\dV{k},
\label{eq:densityElCO}
\end{equation}
where the trace ($\Tr$) indicates a sum of diagonal elements of the resulting
$4\times 4$ matrix. Equivalently, we can write
\begin{align}
	\rho_{\text{e}}(\vect{r}) &= -\int_{\mathcal{K}} \Tr\left[ \phi_{\mu'}(\vect{k};\vect{r})D^{\mu'\mu}(\vect{k})\dg{\phi}_{\mu}(\vect{k};\vect{r}) \right]\dV{k}, \label{eq:densityElBO}\\
	&= -\sum_{\vect{mm}'} \Tr\left[ \chi_{\mu'\vect{m}'}(\vect{r}) D^{\mu'\vect{m}',\mu\vect{m}} \dg{\chi}_{\mu\vect{m}}(\vect{r}) \right]. \label{eq:densityElAO1}
\end{align}
Let us define the 4c \emph{overlap distribution function}
\begin{equation}
\begin{split}
	\Omega_{\mu\vect{m},\mu'\vect{m}'}(\vect{r}) &\equiv \dg{\chi}_{\mu\vect{m}}(\vect{r})\chi_{\mu'\vect{m}'}(\vect{r}) \\
	&= \begin{pmatrix}
		\Omega^{LL}(\vect{r}) & 0_2 \\
		0_2 & \Omega^{SS}(\vect{r})
	\end{pmatrix}_{\mu\vect{m},\mu'\vect{m}'}.
\end{split}
\label{eq:overlapDistDef}
\end{equation}
If we use
\begin{equation}
	\Omega_{\mu\vect{m},\mu'\vect{m}'}(\vect{r}) = \Omega_{\mu\vect{0},\mu'\vect{m}'-\vect{m}}(\vect{r-\vect{m}}),
\label{eq:overlapDistTransl}
\end{equation}
together with the translational invariance of the density matrix
\begin{equation}
	D^{\mu'\vect{m}',\mu\vect{m}} = D^{\mu'\vect{m}'-\vect{m},\mu\vect{0}},
\label{eq:rSpaceDmatTransl}
\end{equation}
then the electron charge density becomes (after changing the summation
variables)
\begin{equation}
	\rho_{\text{e}}(\vect{r}) = -\sum_{\vect{mn}} \Tr\left[ \Omega_{\mu\vect{0},\mu'\vect{m}}(\vect{r}-\vect{n}) D^{\mu'\vect{m},\mu\vect{0}} \right].
\label{eq:densityElAO2}
\end{equation}
We now collect indices $\mu\vect{0},\mu'\vect{m} \equiv u$ and
$\mu'\vect{m},\mu\vect{0} \equiv \bar{u}$, and introduce a shorthand notation
for the trace in real space for an arbitrary operator $A$
\begin{equation}
	A_u D^{\bar{u}} \equiv \sum_{\vect{m}} A_{\mu\vect{0},\mu'\vect{m}} D^{\mu'\vect{m},\mu\vect{0}}.
\label{eq:rSpaceTrace}
\end{equation}
We can then express the total charge density as a sum of nuclear and electronic
contributions
\begin{align}
	\rho(\vect{r}) &= \sum_{\vect{n}} \tilde{\rho}(\vect{r}-\vect{n}), \label{eq:densityTotSum}\\
	\tilde{\rho}(\vect{r}) &= \tilde{\rho}_{\text{n}}(\vect{r}) + \tilde{\rho}_{\text{e}}(\vect{r}), \label{eq:densityTotElNuc}
\end{align}
obtained from the \emph{auxiliary} densities $\tilde{\rho}_{\text{n}}$ and
$\tilde{\rho}_{\text{e}}$ translated from the reference unit cell to the cell
$\vect{n}$.  The auxiliary densities for the reference unit cell are defined as 
\begin{subequations}
\label{eq:densityAuxGroup}
\begin{align}
	\tilde{\rho}_{\text{n}}(\vect{r}) &\equiv \sum_A Z_A \delta(\vect{r}-\vect{A}), \label{eq:densityAuxNuc}\\
	\tilde{\rho}_{\text{e}}(\vect{r}) &\equiv -\Tr\left[ \Omega_u(\vect{r}) D^{\bar{u}} \right],\label{eq:densityAuxEl}
\end{align}
\end{subequations}
where $A$ labels atoms in the reference unit cell, $Z_A$ and $\vect{A}$ being
their charge and position, respectively. Let
\begin{equation}
	N = \sum_{\vect{n}}1
\label{eq:nrUnitCells}
\end{equation}
be the infinite number of unit cells in a crystal and $N_{\text{e}}$ the number
of electrons per unit cell. The electron charge density $\rho_{\text{e}}$ must
integrate to minus the total (infinite) number of electrons, \emph{i.e.}
\begin{equation}
	\int_{\reals^3} \rho_{\text{e}}(\vect{r}) \dV{r} = -NN_{\text{e}}.
\label{eq:integralOfElDensity}
\end{equation}
Hence, we can infer from Eq.~\eqref{eq:densityTotSum} that the auxiliary
electron density $\tilde{\rho}_{\text{e}}$ integrates to minus the number of
electrons per unit cell $N_{\text{e}}$. Moreover, integration of
Eq.~\eqref{eq:densityAuxEl} gives
\begin{equation}
	\Tr\left( S_u D^{\bar{u}} \right) = N_{\text{e}},
\label{eq:traceSDgivesNe}
\end{equation}
where $S_u \equiv S_{\mu\vect{0},\mu'\vect{m}}$ is the 4c overlap matrix from
Eq.~\eqref{eq:rSpaceOverlap4cMat}. Note, however, that whereas the total
electron density $\rho_{\text{e}}$ is a periodic function with the lattice
periodicity, the auxiliary density $\tilde{\rho}_{\text{e}}$ is not periodic.
Nuclear charge densities follow the same arguments. In addition, partitioning
the total density in Eqs.~\eqref{eq:densityTotSum}
and~\eqref{eq:densityTotElNuc} into contributions from individual unit cells
ensures that the lattice sum over $\vect{n}$ is performed in a charge-neutral
manner~\cite{Stolarczyk-IJQC-22-911-1982,Ohnishi-CP-401-152-2012}, \emph{i.e.}
\begin{equation}
	\forall\vect{n}:\quad  \int_{\reals^3} \tilde{\rho}(\vect{r}-\vect{n}) \dV{r} = 0,
\label{eq:densityChargeNeutral}
\end{equation}
provided that there is no excess of positive or negative charge in a unit cell.

\subsection{\label{sec:TimeReversalSymmetry}Time-reversal symmetry}

In the present work, we solve the DKS equation in $\vect{k}$-space
[Eq.~\eqref{eq:DKSinMatrixForm}] by exploiting the time-reversal (TR) symmetry
of the Fock operator. In the absence of a vector potential and in non-magnetic
(non-spin-polarized) systems, TR-symmetric operators attain a special structure
in the so-called \emph{Kramers-restricted}
basis~\cite{Aucar-CPL-232-47-1995,Saue-phdthesis-1996,Saue-MolPhys-91-937-1997,Komorovsky-PRA-94-052104-2016}.
This allows us to reduce the computational and memory resources needed in a
calculation and it also facilitates the interpretation of band structures. Here
we will generalize the concept of a Kramers-restricted GTO basis to reciprocal
space, and explicitly show the structure of the TR-symmetric operators
expressed in this basis.

We start by briefly reviewing the TR operator, which is an antilinear
one-electron operator defined in the 4c realm
as~\cite{Saue-phdthesis-1996,Dyall-92-2007,Komorovsky-PRA-94-052104-2016}
\begin{equation}
	\mathcal{K} = -i\begin{pmatrix}
		\sigma_y & 0_2 \\
		0_2 & \sigma_y
	\end{pmatrix} \mathcal{K}_0,
\label{eq:TROperDef}
\end{equation}
where $\mathcal{K}_0$ denotes complex conjugation. The TR operator satisfies
$\dg{\mathcal{K}}=-\mathcal{K}$ and $\dg{\mathcal{K}}\mathcal{K}=\idmatrix_4$.
An operator $\hat{A}$ is \emph{time-reversal symmetric} iff it commutes with
$\mathcal{K}$ ($[\cdot,\cdot]$ denotes a commutator):
\begin{equation}
	\left[ \hat{A}, \mathcal{K}\right] = 0.
\label{eq:TRSymOperCommut}
\end{equation}
Let us express the TR-symmetric operator $\hat{A}$ in the Kramers-restricted
basis $\left\{\ket{p},\ket{\bar{p}}\right\}$, where
$\ket{\bar{p}}\equiv\mathcal{K}\ket{p}$ denotes the Kramers partner of
$\ket{p}$. If $a\equiv\braket{p|\hat{A}|p}$ and
$b\equiv\braket{p|\hat{A}|\bar{p}}$ label two distinct elements of $A$, then
the remaining 2 elements are given by
\begin{align*}
	\braket{\bar{p}|\hat{A}|p} &= \braket{\mathcal{K}p|\hat{A}|p} = \braket{p|\dg{\mathcal{K}}\hat{A}|p}^{*} =\\
	&= -\braket{p|\mathcal{K}\hat{A}|p}^{*} = -\braket{p|\hat{A}\mathcal{K}|p}^{*} = -b^{*}
\end{align*}
and
\begin{align*}
	\braket{\bar{p}|\hat{A}|\bar{p}} &= \braket{p|\dg{\mathcal{K}}\hat{A}\mathcal{K}|p}^{*} = \braket{p|\hat{A}\dg{\mathcal{K}}\mathcal{K}|p}^{*} =\\
	&= \braket{p|\hat{A}|p}^{*} = a^{*}.
\end{align*}
Hence the matrix representation of the operator $\hat{A}$ has the following
TR-symmetric structure:
\begin{equation}
	A = \begin{pmatrix}
		a & b\\
		-b^{*} & a^{*}
	\end{pmatrix}.
\label{eq:TRSymOperMatrix}
\end{equation}
Note, that the Hermitian adjoint of an antilinear operator involves complex
conjugation of the inner product.

The RKB basis defined in Eq.~\eqref{eq:AObasis4c} is Kramers-restricted in real
space, and can be written
as~\cite{Saue-phdthesis-1996,Saue-MolPhys-91-937-1997}
\begin{equation}
	\chi_{\mu}(\vect{r}) = \begin{pmatrix}
		a & b\\
		-b^{*} & a^{*}
	\end{pmatrix}g_{\mu}(\vect{r}),
\label{eq:RKBisKramersRestr}
\end{equation}
where
\begin{align}
	a &\equiv \begin{pmatrix}
		1 & 0 \\
		0 & \frac{\nabla_z}{2ic} \\
	\end{pmatrix}, & b &\equiv \begin{pmatrix}
		0 & 0 \\
		0 & \frac{\nabla_x}{2ic}-\frac{\nabla_y}{2c} \\
	\end{pmatrix},
\label{eq:RKBisKramersRestrComp}
\end{align}
where we rearranged the $4\times 4$ matrix to emphasize the TR-symmetric
structure of the  basis.  Using the transformation in
Eq.~\eqref{eq:kSpaceBasis}, we obtain the 4c Kramers-restricted Bloch functions
that constitute our basis in $\vect{k}$-space, and which acquire the structure
\begin{equation}
	\phi_{\mu}(\vect{k};\vect{r}) = \begin{pmatrix}
		a(\vect{k};\vect{r}) & b(\vect{k};\vect{r})\\
		-b^{*}(-\vect{k};\vect{r}) & a^{*}(-\vect{k};\vect{r})
	\end{pmatrix}_{\mu},
\label{eq:RKBinKSpace}
\end{equation}
where
\begin{subequations}
\label{eq:RKBinKSpaceGroup}
\begin{align}
	a(\vect{k};\vect{r}) &= \frac{1}{\sqrt{|\mathcal{K}|}} \sum_{\vect{m}} \fourierP{k}{m} a~g_{\mu}(\vect{r}-\vect{m}), \\
	b(\vect{k};\vect{r}) &= \frac{1}{\sqrt{|\mathcal{K}|}} \sum_{\vect{m}} \fourierP{k}{m} b~g_{\mu}(\vect{r}-\vect{m}).
\end{align}
\end{subequations}
As a consequence of the Kramers-restricted basis, the TR-symmetric operator
$\hat{A}$ takes the matrix form of
\begin{equation}
	A_{\mu\vect{0},\mu'\vect{m}} = \begin{pmatrix}
		a & b\\
		-b^{*} & a^{*}
	\end{pmatrix}_{\mu\vect{0},\mu'\vect{m}},
\label{eq:TRSymOperMatrixrSpace}
\end{equation}
in real space, and after the transformation to $\vect{k}$-space
[Eqs.~\eqref{eq:kSpaceGroupFandS}], we have
\begin{equation}
	A_{\mu\mu'}(\vect{k}) = \begin{pmatrix}
		a(\vect{k}) & b(\vect{k})\\
		-b^{*}(-\vect{k}) & a^{*}(-\vect{k})
	\end{pmatrix}_{\mu\mu'},
\label{eq:TRSymOperMatrixkSpace}
\end{equation}
where $a_{\mu\mu'}(\vect{k}) = \sum_{\vect{m}} \fourierP{k}{m}
a_{\mu\vect{0},\mu'\vect{m}}$ (and likewise for $b$).

We now prove two important corollaries of the TR symmetry in our scheme,
namely: 1) that the band energies have a $\vect{k}$-inversion symmetry (as in
the nonrelativistic case); and 2) that the density matrix inherits the TR
structure from the Fock matrix. In addition, it can be shown that a new Fock
matrix constructed from the TR-symmetric density matrix is also TR-symmetric.
This implies that the TR structure is preserved in the self-consistent
procedure, allowing us to impose this structure in the algorithm, significantly
reducing  computational and memory demands in the calculations.  Let us assume
the TR structure in Eq.~\eqref{eq:TRSymOperMatrixkSpace} for the Fock matrix
$F(\vect{k})$, and apply $\mathcal{K}$ from the left to the eigenvalue problem
in Eq.~\eqref{eq:DKSinMatrixForm}
\begin{equation}
	\mathcal{K}F(\vect{k})c(\vect{k}) = \mathcal{K}S(\vect{k})c(\vect{k})\epsilon(\vect{k}).\label{eq:TRSCFkSpace}
\end{equation}
Since $\mathcal{K}$ commutes with the real-space Fock matrix, and trivially
also with the overlap matrix $S(\vect{k})$, it follows that
\begin{subequations}
\label{eq:TRSymOperCommutkGroup}
\begin{align}
	\mathcal{K}F(\vect{k}) &= F(-\vect{k})\mathcal{K}, \label{eq:TRSymOperCommutkFock}\\
	\mathcal{K}S(\vect{k}) &= S(-\vect{k})\mathcal{K}. \label{eq:TRSymOperCommutkOverlap}
\end{align}
\end{subequations}
Flipping $\vect{k}\rightarrow-\vect{k}$ in Eq.~\eqref{eq:TRSCFkSpace} gives
\begin{equation}
	F(\vect{k})\mathcal{K}c(-\vect{k}) = S(\vect{k})\mathcal{K}c(-\vect{k})\epsilon(-\vect{k}).
\label{eq:DKSinMatrixFormTR}
\end{equation}
Because the energies $\epsilon(\vect{k})$ are real, we can infer that
$\left\{c(\vect{k}), \mathcal{K}c(-\vect{k})\right\}$ both are solutions of the
eigenvalue equation Eq.~\eqref{eq:DKSinMatrixForm} with energies
$\left\{\epsilon(\vect{k}), \epsilon(-\vect{k})\right\}$, and thus form a
Kramers pair.
Let us introduce the following notation for the Kramers partners:
\begin{subequations}
\label{eq:TRSymKramersGroup}
\begin{align}
	\bar{c}(\vect{k}) &= \mathcal{K}c(-\vect{k}), \label{eq:TRSymKramersCOs}\\
	\bar{\epsilon}(\vect{k}) &= \epsilon(-\vect{k}), \label{eq:TRSymKramersEnergies}\\
	\bar{f}(\vect{k}) &= f(-\vect{k}), \label{eq:TRSymKramersOccs}
\end{align}
\end{subequations}
where the last equation follows from Eq.~\eqref{eq:FermiDirac}.  In addition,
Eqs.~\eqref{eq:TRSymKramersGroup} imply that the density matrix in reciprocal
space has the TR-symmetric structure of Eq.~\eqref{eq:TRSymOperMatrixkSpace}.
To prove this, we use the block-diagonal structure of the operator
$\mathcal{K}$, and without loss of generality we restrict ourselves to a
$2\times 2$ Fock matrix with solutions
\begin{equation}
	c(\vect{k}) = \begin{pmatrix}
		c^\text{u}(\vect{k}) & \bar{c}^\text{u}(\vect{k}) \\
		c^\text{l}(\vect{k}) & \bar{c}^\text{l}(\vect{k}) \\
	\end{pmatrix},
\label{eq:TRSymKramersCOsMatrix}
\end{equation}
where u and l denote the upper and lower spinor components, respectively.  The
second column is related to the first via the TR operation
Eq.~\eqref{eq:TRSymKramersCOs}, thus $\bar{c}^\text{u}(\vect{k}) =
-c^{\text{l}*}(-\vect{k})$ and $\bar{c}^\text{l}(\vect{k}) =
c^{\text{u}*}(-\vect{k})$.  The density matrix element $D^\text{lu}$ then
satisfies
\begin{align*}
	D^\text{lu}(\vect{k}) = &c^\text{l}(\vect{k})f(\vect{k})c^{\text{u}*}(\vect{k}) + \bar{c}^\text{l}(\vect{k})\bar{f}(\vect{k})\bar{c}^{\text{u}*}(\vect{k}) \\
	= &-\bar{c}^{\text{u}*}(-\vect{k})\bar{f}(-\vect{k})\bar{c}^{\text{l}}(-\vect{k}) \\
	&- c^{\text{u}*}(-\vect{k})f(-\vect{k})c^{\text{l}}(-\vect{k}) \\
	= &-D^{\text{ul}*}(-\vect{k}).
\end{align*}
Similarly $D^\text{ll}(\vect{k}) = D^{uu*}(-\vect{k})$.  It follows that the
real-space elements of the density matrix obtained from
Eq.~\eqref{eq:rSpaceDmatQuadrature} have the TR structure in
Eq.~\eqref{eq:TRSymOperMatrixrSpace}.

\subsection{\label{sec:QuaternionOperators}Quaternion operators}

Owing to the specific structure of TR-symmetric operators, a compact notation
which leads to a very efficient computer implementation can be achieved with
the use of quaternion algebra (or its
isomorphisms)~\cite{Saue-phdthesis-1996,Saue-MolPhys-91-937-1997,Konecny-JCP-149-204104-2018,Repisky-integralPaper}.
This formulation identifies the integrals that are non-redundant and non-zero
when constructing operators in the RKB basis Eq.~\eqref{eq:AObasis4c}, and
allows us to formulate an efficient relativistic algorithm to solve the DKS
equation.  Let $\re x$ and $\im x$ denote the real and imaginary parts of a
complex number $x$, respectively. Then a TR-symmetric matrix $A$ is written as
\begin{equation}
	A = \begin{pmatrix}
		a & b\\
		-b^{*} & a^{*}
	\end{pmatrix} = \sum_{q=0}^3 A^q e_q \equiv A^q e_q,
\label{eq:TRSymOperQuat}
\end{equation}
where
\begin{subequations}
\label{eq:TRSymOperQuatCompGroup}
\begin{align}
	A^0 &= \re a & e_0 &= \hspace{2mm}\idmatrix_2 \equiv 1, \label{eq:TRSymOperQuatComp0}\\
	A^1 &= \im a & e_1 &= i\sigma_z \equiv \check{i},\label{eq:TRSymOperQuatComp1}\\
	A^2 &= \re b & e_2 &= i\sigma_y \equiv \check{j},\label{eq:TRSymOperQuatComp2}\\
	A^3 &= \im b & e_3 &= i\sigma_x \equiv \check{k},\label{eq:TRSymOperQuatComp3}
\end{align}
\end{subequations}
and $\check{i},\check{j},\check{k}$ are fundamental quaternion units obeying
\begin{equation}
	\check{i}^2 = \check{j}^2 = \check{k}^2 = \check{i}\check{j}\check{k} = -1.
\label{eq:quatUnitLaws}
\end{equation}
The Hermitian conjugation of $A$ changes the sign of the three imaginary
components, so that
\begin{align}
	\dg{A} &= \dg{(A^0e_0 + A^1e_1 + A^2e_2 + A^3e_3)} \notag\\
	&= A^{0,T}e_0 - A^{1,T}e_1 - A^{2,T}e_2 - A^{3,T}e_3,
\label{eq:quatHermConj}
\end{align}
where $A^{q,T}$ denotes the transpose of the real matrix $A^q$.
We decompose the TR-symmetric matrices according to
Eq.~\eqref{eq:TRSymOperQuat} and refer to $A^q$ as quaternion components
regardless of whether $e_q$ are interpreted as matrices or quaternion units.
All algebraic manipulations can be performed in an equivalent manner in both
algebras, and it is only a matter of personal preference to select a suitable
representation.  However, we emphasize that encoding complex 4c TR-symmetric
matrices using four real $A^q$ components reduces the number of non-zero terms
by a factor of two, and often reveals important structures of the operators,
facilitating further reductions~\cite{Konecny-JCP-149-204104-2018}.

Matrix elements of a 4c TR-symmetric operator $\hat{A}$ in the basis defined in
Eqs.~\eqref{eq:RKBisKramersRestr} and~\eqref{eq:AObasisPeriodic} are expressed
in real space as
\begin{equation}
	A_{\mu\vect{0},\mu'\vect{m}} = A^q_{\mu\vect{0},\mu'\vect{m}} e_q,
\label{eq:TRSymOperQuatBasis}
\end{equation}
where $A^q_{\mu\vect{0},\mu'\vect{m}}$ are $2\times 2$ real matrices:
\begin{equation}
	A^q_{\mu\vect{0},\mu'\vect{m}} = \begin{pmatrix}
		A^{LL,q} & A^{LS,q} \\
		A^{SL,q} & A^{SS,q}
	\end{pmatrix}_{\mu\vect{0},\mu'\vect{m}}.
\label{eq:TRSymOperQuatBasisComp}
\end{equation}
Reciprocal-space quaternion components of $A$ are defined by the Fourier series
\begin{equation}
	A^q_{\mu\mu'}(\vect{k}) = \sum_{\vect{m}}\fourierP{k}{m} A^q_{\mu\vect{0},\mu'\vect{m}},
\label{eq:TRSymOperQuatkSpaceComp}
\end{equation}
and form a quaternion (dropping the $\mu\mu'$ indices)
\begin{equation}
	A(\vect{k}) = A^q(\vect{k})e_q,
\label{eq:TRSymOperQuatkSpace}
\end{equation}
with complex-valued components $A^q(\vect{k})$.

During the self-consistent procedure, we exchange the quaternion form of the
Fock matrix with its complex form Eq.~\eqref{eq:TRSymOperMatrixkSpace}, and
vice versa. Whereas the quaternion form is more beneficial in real space to
facilitate the integral evaluation when assembling the Fock matrix, the matrix
form is inevitable in the diagonalization step of the procedure. Additionally,
if we establish a direct connection between these forms in reciprocal space, we
avoid unnecessary computations of the Fourier series, because there are
considerably fewer nonzero quaternion components than complex matrix elements.
Therefore, we use the definitions in Eqs.~\eqref{eq:TRSymOperQuatCompGroup}
together with Eq.~\eqref{eq:TRSymOperQuatkSpace} to compose a complex matrix
\begin{equation}
	A(\vect{k}) \equiv \begin{pmatrix}
		\phantom{-}A^0(\vect{k})+iA^1(\vect{k}) & A^2(\vect{k})+iA^3(\vect{k}) \\
		-A^2(\vect{k})+iA^3(\vect{k}) & A^0(\vect{k})-iA^1(\vect{k})
	\end{pmatrix}.
\label{eq:TRSymOperQuat2Cmat}
\end{equation}
This matrix is consistent with Eq.~\eqref{eq:TRSymOperMatrixkSpace}, because
the definition of the reciprocal-space quaternion components
[Eq.~\eqref{eq:TRSymOperQuatkSpaceComp}] implies
\begin{equation}
	A^{q*}(\vect{k}) = A^q(-\vect{k}).
\label{eq:TRSymOperQuatkSpaceCompCC}
\end{equation}
Inverting this process allows us to map a complex matrix
\begin{equation*}
	A(\vect{k}) = \begin{pmatrix}
		a(\vect{k}) & b(\vect{k}) \\
		c(\vect{k}) & d(\vect{k})
	\end{pmatrix}
\end{equation*}
with assumed TR symmetry [$c(\vect{k}) = -b^*(-\vect{k})$, $d(\vect{k}) =
a^*(-\vect{k})$] to a quaternion with complex components given by
\begin{subequations}
\label{eq:TRSymOperCmat2QuatGroup}
\begin{align}
	A^0(\vect{k}) &= \frac{1}{2 }\left[ a(\vect{k})+d(\vect{k}) \right], \\
	A^1(\vect{k}) &= \frac{1}{2i}\left[ a(\vect{k})-d(\vect{k}) \right], \\
	A^2(\vect{k}) &= \frac{1}{2 }\left[ b(\vect{k})-c(\vect{k}) \right], \\
	A^3(\vect{k}) &= \frac{1}{2i}\left[ b(\vect{k})+c(\vect{k}) \right].
\end{align}
\end{subequations}
For $\vect{k}=\vect{0}$ quaternion components, $A^q(\vect{0})$ are real, and
Eqs.~\eqref{eq:TRSymOperCmat2QuatGroup} coincide with the definitions in
Eqs.~\eqref{eq:TRSymOperQuatCompGroup}.

We now rewrite all operators in Eqs.~\eqref{eq:rSpace2cGroup} that enter the
DKS equation in the language of quaternions. Scalar operators $\mathcal{S},
\mathcal{T}$, and $\mathcal{V}^{LL}$ have a trivial structure in the spin
space, therefore their corresponding quaternions have nonzero real part (0-th
component) and zero imaginary part. On the other hand, the operator
$\mathcal{V}^{SS}$ contains Pauli matrices, and thus is a general quaternion
$\mathcal{V}^{SS} = \mathcal{V}^{SS,q}e_q$. The Fock matrix in
Eq.~\eqref{eq:rSpaceFock4cMat} can then be expressed as (omitting
$\mu\vect{0},\mu'\vect{m}$ indices for clarity)
\begin{equation}
	F = \begin{pmatrix}
		\mathcal{V}^{LL,0} & \mathcal{T}^0 \\
		\mathcal{T}^0 & \frac{1}{4c^2}\mathcal{V}^{SS,0}-\mathcal{T}^0
	\end{pmatrix}e_0 + \begin{pmatrix}
		0 & 0 \\
		0 & \frac{1}{4c^2}\mathcal{V}^{SS,i}
	\end{pmatrix}e_i,
\label{eq:rSpaceFock4cQuat}
\end{equation}
for $i=1,2,3$. It is convenient to rewrite the potential $V$ in terms of the 4c
overlap distribution $\Omega$ defined in Eq.~\eqref{eq:overlapDistDef}. We
accomplish this by rewriting Eq.~\eqref{eq:rSpace2cWMat} as
\begin{equation*}
	\mathcal{V}^{SS}_{\mu\vect{0},\mu'\vect{m}} = \int_{\reals^3} \dg{\left[(\vect{\sigma}\cdot\vect{p})g_{\mu\vect{0}}(\vect{r})\right]} \left[(\vect{\sigma}\cdot\vect{p}) g_{\mu'\vect{m}}(\vect{r})\right] V(\vect{r}) \dV{r}.
\end{equation*}
The small-component overlap distribution is a product of small-component basis
functions [Eq.~\eqref{eq:AObasis4cSS}], so
\begin{equation}
	\Omega^{SS}_{\mu\vect{0},\mu'\vect{m}} = \frac{1}{4c^2}\dg{\left[ (\vect{\sigma}\cdot\vect{p})g_{\mu\vect{0}} \right]} \left[ (\vect{\sigma}\cdot\vect{p})g_{\mu'\vect{m}} \right].
\label{eq:Omega4cSSExplicit}
\end{equation}
The potential therefore becomes
\begin{subequations}
\label{eq:VmatOmegaGroup}
\begin{align}
	\mathcal{V}^{LL}_u &= \int_{\reals^3} \Omega^{LL}_u(\vect{r})V(\vect{r}) \dV{r}, \label{eq:VmatOmegaLL}\\
	\frac{1}{4c^2}\mathcal{V}^{SS}_u &= \int_{\reals^3} \Omega^{SS}_u(\vect{r})V(\vect{r}) \dV{r}, \label{eq:VmatOmegaSS}
\end{align}
\end{subequations}
where $u \equiv \mu\vect{0},\mu'\vect{m}$, and the overlap distributions are
quaternions
\begin{subequations}
\label{eq:Omega4cQuatGroup}
\begin{align}
	\Omega^{LL}_u(\vect{r}) &= \Omega^{LL,0}_u(\vect{r}) e_0, \label{eq:Omega4cQuatLL}\\
	\Omega^{SS}_u(\vect{r}) &= \Omega^{SS,q}_u(\vect{r}) e_q. \label{eq:Omega4cQuatSS}
\end{align}
\end{subequations}
Explicit forms of the quaternion components of $\Omega^{SS}$ can be identified
if we apply the multiplication rule for the Pauli matrices to
Eq.~\eqref{eq:Omega4cSSExplicit}, \emph{i.e.}
\begin{equation}
\begin{split}
	\Omega^{SS}_{\mu\vect{0},\mu'\vect{m}} &= \frac{1}{4c^2} \dg{\left(\vect{\nabla}g_{\mu\vect{0}}\right)} \cdot \left(\vect{\nabla}g_{\mu'\vect{m}}\right)\idmatrix_2 \\
	&+ \frac{1}{4c^2} \dg{\left(\vect{\nabla}g_{\mu\vect{0}}\right)} \times \left(\vect{\nabla}g_{\mu'\vect{m}}\right) \cdot i\vect{\sigma}.
\end{split}
\label{eq:Omega4cSSExplicit2}
\end{equation}
This analysis shows that in order to build 4c complex matrices for the Coulomb
and exchange--correlation operators, it is sufficient to evaluate integrals in
Eqs.~\eqref{eq:VmatOmegaGroup} for five components of the overlap distribution --
one for the $LL$ sector, and four for the $SS$ sector. The $\vect{k}$-space matrix
is then obtained by computing the Fourier series of these five components
[Eq.~\eqref{eq:TRSymOperQuatkSpaceComp}] and arranging them according to
Eq.~\eqref{eq:TRSymOperQuat2Cmat}. Moreover, one can obtain a spin-free form of
the DKS equation in solids by omitting the imaginary quaternion terms that are
associated with the spin--orbit interaction, in analogy to the procedure
proposed by Dyall for molecules\cite{Dyall-JCP-100-2118-1994}.

We conclude this section by employing the quaternion formalism to express
expectation values (traces with the density matrix) of TR-symmetric operators
appearing in the DKS equation. Suppose a matrix $A$ has the same structure as
the potential $V$, \emph{i.e.} does not couple the large and small components
of the wave function, and its $LL$ quaternion has zero imaginary part. Its
trace with a density matrix $D$, as defined in Eq.~\eqref{eq:rSpaceTrace}, is
obtained by using the traceless property of the Pauli matrices as 
\begin{align}
	\Tr \left[A_uD^{\bar{u}}\right] = &\Tr\left[\begin{pmatrix}
		A^{LL} & 0_2 \\
		0_2 & A^{SS}
	\end{pmatrix}_u \begin{pmatrix}
		D_{LL} & D_{LS} \\
		D_{SL} & D_{SS}
	\end{pmatrix}^{\bar{u}}\right] \notag\\
	= & 2\left(A^{LL,0}_uD_{LL,0}^{\bar{u}} + A^{SS,0}_uD_{SS,0}^{\bar{u}}\right. \label{eq:traceTRsymAD}\\
	&- \left. A^{SS,i}_uD_{SS,i}^{\bar{u}}\right), \notag
\end{align}
implicitly summing over $u$ and $i=1,2,3$. Note that despite the general
TR-symmetric structure of the density matrix, only its corresponding five elements
are required to evaluate the trace. Equation~\eqref{eq:traceTRsymAD} also holds for
the electron density in Eq.~\eqref{eq:densityAuxEl} when substituting
$A_u\rightarrow\Omega_u(\vect{r})$.  The kinetic energy operator $T$
[Eq.~\eqref{eq:rSpaceTmatAndVmat4comp}] has a different structure than the
potential $V$. We evaluate its trace with the density matrix to compute the
kinetic energy per unit cell as
\begin{equation*}
	\frac{E_\text{k}}{N} = \Tr \left[T_uD^{\bar{u}}\right] = \Tr\left[\begin{pmatrix}
		0_2 & \mathcal{T} \\
		\mathcal{T} & -\mathcal{T}
	\end{pmatrix}_u \begin{pmatrix}
		D_{LL} & D_{LS} \\
		D_{SL} & D_{SS}
	\end{pmatrix}^{\bar{u}}\right].
\end{equation*}
It follows that
\begin{equation}
	\frac{E_\text{k}}{N} = 2\mathcal{T}_u^0\left( D_{SL,0}^{\bar{u}} + D_{LS,0}^{\bar{u}} - D_{SS,0}^{\bar{u}} \right).
\label{eq:energyKinetic}
\end{equation}

\subsection{\label{sec:CoulombPotentialAndEnergy}Coulomb potential and energy}

Using the auxiliary charge density $\tilde{\rho}$ from
Eq.~\eqref{eq:densityTotSum}, we can express the Coulomb contribution $J$ to
the Fock matrix in Eq.~\eqref{eq:rSpaceFock4cMatParts}
\begin{equation}
	J(\vect{r}) = -\int_{\reals^3} \frac{\rho(\vect{r}')\dV{r}'}{|\vect{r}-\vect{r}'|},
\label{eq:CoulombOperatorDef}
\end{equation}
as
\begin{equation}
	J(\vect{r}) = -\sum_{\vect{n}}\int_{\reals^3} \frac{\tilde{\rho}(\vect{r}')\dV{r}'}{|\vect{r}-\vect{r}'-\vect{n}|}.
\label{eq:CoulombOperatorAuxDens}
\end{equation}
We see that the Coulomb potential is a periodic function with the lattice
periodicity, given that the lattice sum over $\vect{n}$ runs over the entire
infinite lattice. Any truncation of this sum (for instance, for numerical
purposes) will violate the translational symmetry. We express the
non-equivalent matrix elements of $J$ in the real-space basis defined in
Eqs.~\eqref{eq:AObasisPeriodic} and~\eqref{eq:AObasis4c} as
\begin{equation*}
	J_{\mu\vect{0},\mu'\vect{m}} = \int_{\reals^3} \dg{\chi}_{\mu\vect{0}}(\vect{r}) J(\vect{r})\chi_{\mu'\vect{m}}(\vect{r}) \dV{r}.
\end{equation*}
Since the Coulomb potential $J(\vect{r})$ is diagonal in the $4\times 4$
bispinor space, it follows that
\begin{equation}
\begin{split}
	J_u &= \int_{\reals^3} \Omega_u(\vect{r})J(\vect{r}) \dV{r} \\
	&= -\sum_{\vect{n}} \int_{\reals^3\times\reals^3} \frac{\Omega_u(\vect{r}_1)\tilde{\rho}(\vect{r}_2)}{|\vect{r}_1-\vect{r}_2-\vect{n}|} \dV{r}_1\dV{r}_2,
\end{split}
\label{eq:CoulombMatrixOmega}
\end{equation}
where $u\equiv\mu\vect{0},\mu'\vect{m}$, and $\Omega$ is the 4c overlap
distribution defined in Eqs.~\eqref{eq:overlapDistDef}
and~\eqref{eq:Omega4cQuatGroup}. Substituting the nuclear and electronic
auxiliary densities [Eqs.~\eqref{eq:densityAuxGroup}], we obtain
\begin{subequations}
\label{eq:CoulombMatrixFinalGroup}
\begin{align}
	J_u &= \sum_{\vect{n}} \big[ J^{\text{n}}_u(\vect{n})+J^{\text{e}}_u(\vect{n}) \big], \label{eq:CoulombMatrixFinalTot}\\
	J^{\text{n}}_u(\vect{n}) &= \sum_A\int_{\reals^3} \frac{-Z_A\Omega_u(\vect{r})}{|\vect{r}-\vect{A}-\vect{n}|} \dV{r}, \label{eq:CoulombMatrixFinalNuc}\\
	J^{\text{e}}_u(\vect{n}) &= \int_{\reals^3\times\reals^3} \frac{\Omega_u(\vect{r}_1)\Tr\left[ \Omega_v(\vect{r}_2)D^{\bar{v}} \right]}{|\vect{r}_1-\vect{r}_2-\vect{n}|} \dV{r}_1\dV{r}_2. \label{eq:CoulombMatrixFinalEl}
\end{align}
\end{subequations}
Note that in Eq.~\eqref{eq:CoulombMatrixFinalEl}, the sum over
$v\equiv\nu\vect{0},\nu'\vect{n}'$ is implied. This sum over $v$ together with
the lattice sum over $\vect{n}$ in Eq.~\eqref{eq:CoulombMatrixFinalTot} must be
computed for each $u\equiv\mu\vect{0},\mu'\vect{m}$, making this term the most
computationally expensive to evaluate.

The expression for the Coulomb energy in a periodic system can be obtained in a
similar manner. Inserting the auxiliary density to
\begin{equation}
	E_{\text{C}} = \frac{1}{2} \int_{\reals^3\times\reals^3} \frac{\rho(\vect{r}_1)\rho(\vect{r}_2)}{|\vect{r}_1-\vect{r}_2|} \dV{r}_1\dV{r}_2,
\label{eq:CoulombEnergyDef}
\end{equation}
gives
\begin{equation}
	\frac{E_{\text{C}}}{N} = \frac{1}{2} \sum_{\vect{n}}\int_{\reals^3\times\reals^3} \frac{\tilde{\rho}(\vect{r}_1)\tilde{\rho}(\vect{r}_2)}{|\vect{r}_1-\vect{r}_2-\vect{n}|} \dV{r}_1\dV{r}_2.
\label{eq:CoulombEnergyAuxDens}
\end{equation}
If we divide the density into nuclear and electron contributions, and use the
definitions in Eqs.~\eqref{eq:CoulombMatrixFinalNuc}
and~\eqref{eq:CoulombMatrixFinalEl}, we obtain
\begin{equation}
	\frac{E_{\text{C}}}{N} = \frac{1}{2}\sum_{\vect{n}} E_{\text{nn}}(\vect{n}) + 2\Tr\left[ J^{\text{n}}_u(\vect{n})D^{\bar{u}}\right] + \Tr\left[ J^{\text{e}}_u(\vect{n})D^{\bar{u}}\right],
\label{eq:CoulombEnergyFinal}
\end{equation}
where
\begin{equation}
	E_{\text{nn}}(\vect{n}) = \overline{\sum_{AB}} \frac{Z_AZ_B}{\left| \vect{A} - \vect{B} - \vect{n} \right|}
\label{eq:CoulombEnergyNucNuc}
\end{equation}
is the nuclear--nuclear repulsion energy, and the bar over the sum indicates
that the divergent self-interaction terms are excluded. The traces of the 4c
matrices $J^{\text{n}}_u(\vect{n})$ and $J^{\text{e}}_u(\vect{n})$ with the
density matrix are evaluated using Eq.~\eqref{eq:traceTRsymAD}. In
Eq.~\eqref{eq:CoulombEnergyFinal}, we grouped the electron--nuclear and
nuclear--electron terms together --- this is only possible if
$\sum_{\vect{n}}J^{\text{n}}_u(-\vect{n}) =
\sum_{\vect{n}}J^{\text{n}}_u(\vect{n})$, so the lattice sum must contain both
the $\vect{n}$ and $-\vect{n}$ unit cells for each $\vect{n}$. This is true for
the infinite lattice sum, but should be taken into account when designing
approximations to the lattice sum.

\subsection{\label{sec:TreatmentOfElectrostaticLatticeSums}Treatment of electrostatic lattice sums}

A complication that emerges when studying periodic systems is the evaluation of
the electrostatic lattice sums $\sum_{\vect{n}}$ that appear in the Coulomb
potential [Eq.~\eqref{eq:CoulombMatrixFinalTot}] and the Coulomb energy
[Eq.~\eqref{eq:CoulombEnergyFinal}]. The difficulty originates in the
long-range nature of the electrostatic Coulomb interaction, and manifests
itself in two ways. One issue is the question of the convergence itself. The
lattice sums of individual electronic and nuclear contributions to the
potential and energy are divergent, hence they must be treated in a
charge-neutral manner, such as in Eqs.~\eqref{eq:CoulombMatrixFinalTot}
and~\eqref{eq:CoulombEnergyFinal}.  Assuming that the unit cell is electrically
neutral, the charge-neutral lattice sums are convergent. Unfortunately, their
convergence is often only conditional, and therefore the result is not
determined uniquely unless physical arguments are incorporated. In such cases,
the results can be shown to depend both on the choice of the unit cell
shape~\cite{Piela-CPL-86-195-1982}, as well as on the implemented summation
technique~\cite{Redlack-JPCS-36-73-1975}.  The convergence problems were
rigorously investigated by de Leeuw, Perram, and
Smith~\cite{Leeuw-PRSA-373-27-1980}, who introduced convergence factors to
enforce absolute converge on the lattice sums.  The second complication is the
very slow convergence of the sums. Even if the sum is absolutely convergent,
imprudent truncation of the sums severely distorts the potential and breaks its
translational invariance. To enable the evaluation of the electrostatic
potential and energy, the Coulomb operator is expanded in a spherical multipole
expansion [Eq.~\eqref{eq:SMECompact} with $\vect{P}=\vect{0}$ and
$\vect{Q}=\vect{n}$]
\begin{equation}
	\frac{1}{|\vect{r}_1-\vect{r}_2-\vect{n}|} = R^T(\vect{r}_1)\Theta(\vect{n})R(\vect{r}_2),
\label{eq:SMECompactOperator}
\end{equation}
where $R$ is the vector of scaled regular solid harmonics, and $\Theta$ is the
interaction tensor, defined in the work of Watson \emph{et
al.}~\cite{Watson-JCP-121-2915-2004} (see also Ref.~\cite{Helgaker-405-2000}
and Appendix~\ref{app:SphericalMultipoleExpansion}).  The Coulomb problem is
then reduced to the computation of the lattice sum of the spherical interaction
tensors. Nijboer and De Wette proposed a universal method for computing such
lattice sums~\cite{Nijboer-Physica-23-309-1957}. Their approach is based on an
Ewald-like partitioning of the sums into terms that converge rapidly in direct
space, and terms that converge rapidly in reciprocal space. In this work, we
follow a scheme that employs a renormalization identity, first introduced by
Berman and Greengard~\cite{Berman-JMP-35-6036-1994}, and then later
reformulated by Kudin and Scuseria~\cite{Kudin-JCP-121-2886-2004}.  Contrary to
the approach of Kudin and Scuseria, we factor out the sum of the interaction
tensors $\Theta(\vect{n})$, as shown later in this section.  Because the sum of
the interaction tensors only depends on the lattice parameters, we
pre-calculate it before proceeding to the solution of the DKS equations.

We now apply the spherical multipole expansion in
Eq.~\eqref{eq:SMECompactOperator} to derive expressions for the Coulomb
potential and energy. First we split the infinite lattice sum over $\vect{n}$
in Sec.~\ref{sec:CoulombPotentialAndEnergy}
\begin{equation}
	\sum_{\vect{n}} = \sum_{\vect{n}\in\text{NF}} + \sum_{\vect{n}\in\text{FF}},
\label{eq:latticeSumNFandFFsplitting}
\end{equation}
where NF is the near-field and FF is the far-field of the reference unit cell
$\vect{n}=\vect{0}$. The FF is constructed to contain all unit cells for which
a universal multipole expansion in Eq.~\eqref{eq:SMECompactOperator} centered
in $\vect{n}=\vect{0}$ produces a globally valid approximation to the integrals
in Eqs.~\eqref{eq:CoulombMatrixFinalGroup}. A remaining finite array of unit
cells constitutes the NF. Our partitioning scheme is similar to those discussed
in previous
studies~\cite{Lazarski-JCTC-11-3029-2015,White-CPL-230-8-1994,Watson-JCP-121-2915-2004}.
Inserting the multipole expansion in Eq.~\eqref{eq:SMECompactOperator} into
Eqs.~\eqref{eq:CoulombMatrixOmega} and~\eqref{eq:CoulombEnergyAuxDens} gives
the corresponding contributions to the far-field potential and energy
\begin{align}
	J_u^{\text{FF}} &= q_u^T \Lambda Q, \label{eq:FFCoulombMatrixFinal}\\
	\frac{E^{\text{FF}}_{\text{C}}}{N} &= \frac{1}{2}Q^T\Lambda Q. \label{eq:FFCoulombEnergyFinal}
\end{align}
We have here defined the lattice sum of interaction tensors
\begin{equation}
	\Lambda_{lm,jk} \equiv \sum_{\vect{n}\in\text{FF}} \Theta_{lm,jk}(\vect{n}),
\label{eq:latSumOfIntTens}
\end{equation}
elements of the 4c electronic multipole moment operator
\begin{equation}
	q_u^{lm} \equiv - \int_{\reals^3} \Omega_u(\vect{r})R^{lm}(\vect{r}) \dV{r},
\label{eq:multMomInegral}
\end{equation}
and the total multipole moments of the reference unit cell
\begin{equation}
	Q^{lm} = \int_{\reals^3} \tilde{\rho}(\vect{r})R^{lm}(\vect{r}) \dV{r}.
\label{eq:multMomTot1}
\end{equation}
Inserting the definition of the auxiliary density from
Eqs.~\eqref{eq:densityTotElNuc} and~\eqref{eq:densityAuxGroup} to
Eq.~\eqref{eq:multMomTot1} gives a more convenient expression for the total
multipole moments
\begin{equation}
	Q^{lm} = \sum_A Z_AR^{lm}(\vect{A}) + \Tr\left[ q_u^{lm}D^{\bar{u}} \right],
\label{eq:multMomTot2}
\end{equation}
where we implied the summation over $u$ as defined in
Eq.~\eqref{eq:rSpaceTrace}.  The trace of $q^{lm}_u$ with the density matrix is
computed as in Eq.~\eqref{eq:traceTRsymAD}.  Notice that the total charge
$Q^{00}=0$, because $R^{00}=1$, $q^{00}_u = -S_u$, and $\Tr\left[
  S_uD^{\bar{u}} \right]=N_e$. Furthermore, $Q^{1m}$ is the total (electric $+$
nuclear) dipole moment, which is gauge origin independent. To summarize, by
employing the multipole expansion we accomplished two tasks: We isolated the
slow-converging lattice sum $\sum_{\vect{n}}$, facilitating its subsequent
computation, and we factorized the complicated six-dimensional two-electron
integrals in Eq.~\eqref{eq:CoulombMatrixFinalEl} into a product of simpler
three-dimensional one-electron integrals [Eq.~\eqref{eq:multMomInegral}]. In
this way, we obtained a very efficient scheme to incorporate the potential
generated by the infinite lattice.

Analysis of the multipole expansion reveals that the problem of the conditional
convergence of the Coulomb series can be attributed to non-zero unit cell
dipole and quadrupole moments~\cite{Leeuw-PRSA-373-27-1980}. In fact, the
three-dimensional lattice sums of the $\Theta_{1m,00}$ and $\Theta_{00,1k}$
elements of the interaction tensor that enter the far-field potential
[Eq.~\eqref{eq:FFCoulombEnergyFinal}] are divergent. To rectify this, we
introduce fictitious point charges at unit cell face centers, as was done in
previous studies~\cite{Stolarczyk-IJQC-22-911-1982,Kudin-CPL-283-61-1998}. For
each of the three periodic dimensions $i=1,2,3$, two charges $\pm z^i$ are
placed at opposing walls $\pm\frac{\vect{a}_i}{2}$ for each unit cell. This
procedure guarantees that the unit cell remains charge neutral. Furthermore,
every unit cell wall is shared by 2 unit cells, and thus contains 2 fictitious
charges with opposite signs, canceling each other. Note that this scheme is
valid for arbitrary unit-cell geometries. The values $z^i$ are determined so
that they eliminate the unit cell dipole moment $\vect{\mu}$, and they are
obtained by solving a linear system of equations
\begin{equation}
	z^i\vect{a}_i = -\vect{\mu}.
\label{eq:fictChargeCompensation}
\end{equation}
To understand how the inclusion of fictitious charges resolves the problem of
the conditional convergence, let us enclose a crystal sample in a finite
volume, and examine the limit of the (finite) lattice sum over unit cells
inside the volume as the volume approaches infinity. The lattice sum in the
Coulomb potential and energy can be shown to contain surface-dependent terms
that are linear and quadratic in the position, and hence break the periodicity
of the potential~\cite{Redlack-JPCS-36-73-1975,Kantorovich-JPCM-11-6159-1999}.
These terms do not vanish in the limit of the infinite volume, and thus the
limit gives different results for different volume shapes. The fictitious
charges included as described above only cancel inside the volume, not on its
surface, and serve to compensate the ambiguous linear (charge--dipole) surface
terms in the potential. Quadratic (charge--quadrupole) surface terms could be
eliminated similarly, but because they simply shift the potential by a
constant, they are ignored in this work. Such shifts affect absolute band
energies, but do not alter the total energy or the band gaps.

We conclude this section by adapting the renormalization procedure of Kudin and
Scuseria~\cite{Kudin-JCP-121-2886-2004} to the evaluation of the lattice sum in
Eq.~\eqref{eq:latSumOfIntTens}. Instead of a direct calculation, the sum
$\Lambda$ is obtained as a limit
\begin{equation}
	\Lambda = \lim_{t\rightarrow\infty} \Lambda^t.
\label{eq:latSumAsLimit}
\end{equation}
$\Lambda^t$ are partial sums that are computed by iterating the recurrence
equation
\begin{equation}
	\Lambda^{t+1} = \Lambda^1 + \mathcal{U}(\Lambda^t)\mathcal{W},
\label{eq:latSumRenormEq}
\end{equation}
where
\begin{equation}
	\mathcal{U}(\Lambda^t_{lm,jk}) = \frac{1}{3^{l+j+1}}\Lambda^t_{lm,jk}
\label{eq:latSumScalingOp}
\end{equation}
is the scaling operator, and
\begin{equation}
	\mathcal{W} = \sum_{\mu^1\ldots \mu^d=-1}^1 W(\mu^i\vect{a}_i)
\label{eq:latSumTranslSum}
\end{equation}
is a matrix consisting of a sum of translation tensors $W$ defined in
Appendix~\ref{app:SphericalMultipoleExpansion}. The recurrence scheme is
initiated by
\begin{equation}
	\Lambda^1 = \sum_{\vect{n}\in\text{FF}_1}\Theta(\vect{n}) \equiv \sum_{n^1\ldots n^d\in\text{FF}_1} \Theta(n^i\vect{a}_i),
\label{eq:latSumInitial}
\end{equation}
where FF$_1$ contains all unit cells that are in the far-field of the central
reference unit cell, but are in the near-field of the supercell composed of the
original near-field. To illustrate this, let the near-field supercell be a
block (in crystallographic coordinates) consisting of unit cells with indices
$n^i=-N_i,\ldots,N_i$ for each of the periodic dimensions $i=1,\ldots,d$. Thus
the total number of unit cells in such a block is $\prod_{i=1}^d (2N_i+1)$.
Then
\begin{equation}
	\text{FF}_1 = \left\{ (n^1\ldots n^d)\in\mathbb{Z}^d; 1\leq \max_{i=1\ldots d}\left(\frac{|n^i|-1}{N_i}\right) \leq 3 \right\}.
\label{eq:FF0}
\end{equation}
In contrast to a naive term-by-term summation, the recurrence formula
[Eq.~\eqref{eq:latSumRenormEq}] converges rapidly to its limit, and in practice
only a few iterations are needed. We provide a formal derivation of
Eq.~\eqref{eq:latSumRenormEq} in
Appendix~\ref{app:LatticeSumOfInteractionTensors}.

\subsection{\label{sec:ExchangeCorrelationContribution}Exchange--correlation contribution}

We here derive the exchange--correlation (XC) contribution to the Fock operator
and the energy of periodic systems. We assume the non-relativistic generalized
gradient approximation (GGA) for the XC energy
functional~\cite{Becke-PRA-38-3098-1988,Perdew-PRB-46-6671-1992}.  Within the
Kramers-restricted (closed shell) framework, a GGA-type XC functional is
expressed as
\begin{equation}
	E_\text{XC}\left[n,\vect{\nabla}n\right] \equiv E_{\text{XC}} = \int_{\reals^3} \epsilon_{\text{XC}}(\vect{r})\dV{r},
\label{eq:XCEnergyTot}
\end{equation}
where $\epsilon_{\text{XC}}(\vect{r}) \equiv
\epsilon_{\text{XC}}\left[n,\vect{\nabla}n\right](\vect{r})$ is the XC energy
density, and $n(\vect{r})$ is the total electron probability density obtained
from the electron charge density in Eq.~\eqref{eq:densityElAO2} as $n(\vect{r})
\equiv -\rho_\text{e}(\vect{r})$.  For periodic systems, the integration over
$\reals^3$ can be limited to an integration over the central reference unit
cell, because the electron density is a periodic function with the lattice
periodicity, and consequently $\epsilon_\text{XC}(\vect{r}+\vect{m}) =
\epsilon_\text{XC}(\vect{r})$. Letting $\mathcal{C}_{\vect{m}}$ denote the unit
cell positioned at the lattice point $\vect{m}$, we obtain
\begin{align*}
	E_\text{XC} &= \sum_{\vect{m}} \int_{\mathcal{C}_{\vect{m}}} \epsilon_{\text{XC}}(\vect{r})\dV{r} = \sum_{\vect{m}} \int_{\mathcal{C}_{\vect{0}}} \epsilon_{\text{XC}}(\vect{r}+\vect{m})\dV{r} \\
	&= \sum_{\vect{m}} \int_{\mathcal{C}_{\vect{0}}} \epsilon_{\text{XC}}(\vect{r})\dV{r} = N \int_{\mathcal{C}_{\vect{0}}} \epsilon_{\text{XC}}(\vect{r})\dV{r},
\end{align*}
where $N$ is the total number of unit cells. Therefore, the XC energy per unit
cell is
\begin{equation}
	\frac{E_\text{XC}}{N} = \int_{\mathcal{C}_{\vect{0}}} \epsilon_{\text{XC}}(\vect{r})\dV{r}.
\label{eq:XCEnergyUC}
\end{equation}

The XC functional has a complicated dependence on the electron density, and the
integral in Eq.~\eqref{eq:XCEnergyUC} must therefore be integrated numerically.
Because the integrand $\epsilon_\text{XC}$ is a highly inhomogeneous function
in real space containing cusps, a robust numerical technique is needed. In this
work we follow the integration scheme developed by Towler \emph{et
al.}~\cite{Towler-CPC-98-181-1996}, which is an extension of Becke's atomic
partitioning method~\cite{Becke-JCP-88-2547-1988} to periodic systems. Towler
\emph{et al.} introduced a weight function $w_A(\vect{r})$ for each atom $A$ in
the reference unit cell, and define it for all other unit cells
$\mathcal{C}_{\vect{m}}$ using translations:
\begin{equation}
	w_{A\vect{m}}(\vect{r}) \equiv w_A(\vect{r}-\vect{m}).
\label{eq:XCWeightsTranslated}
\end{equation}
The weight functions are constructed to be normalized to unity for each point
$\vect{r}$, \emph{i.e.}
\begin{equation}
	\sum_{A\vect{m}} w_{A\vect{m}}(\vect{r}) = 1.
\label{eq:XCWeightsNormalization}
\end{equation}
The detailed process of forming the weight functions can be found in
Refs.~\cite{Becke-JCP-88-2547-1988,Towler-CPC-98-181-1996}. Inserting the
weights into Eq.~\eqref{eq:XCEnergyUC} gives
\begin{align*}
	\frac{E_\text{XC}}{N} &= \int_{\mathcal{C}_{\vect{0}}} \epsilon_{\text{XC}}(\vect{r}) \sum_{A\vect{m}} w_A(\vect{r}-\vect{m}) \dV{r}\\
	&= \sum_{A\vect{m}} \int_{\mathcal{C}_{-\vect{m}}} \epsilon_{\text{XC}}(\vect{r})w_A(\vect{r}) \dV{r}.
\end{align*}
It follows, that
\begin{equation}
	\frac{E_\text{XC}}{N} = \sum_A \int_{\reals^3} \epsilon_{\text{XC}}(\vect{r})w_A(\vect{r}) \dV{r}.
\label{eq:XCEnergyFinal}
\end{equation}
For a discrete set of grid points $\vect{g}$, the integral is replaced by a
weighted sum
\begin{equation}
	\frac{E_\text{XC}}{N} \rightarrow \sum_{\vect{g}} \epsilon_{\text{XC}}(\vect{g})w(\vect{g}),
\label{eq:XCEnergyFinalGrid}
\end{equation}
where the sum is over an integration grid composed of the joined atomic grids
and, similarly, the weights $w(\vect{g})$ contain all atomic weights
$w_A(\vect{g})$.

The XC potential is defined as the functional derivative of the XC energy:
\begin{equation}
	V^\text{XC}(\vect{r}) = \frac{\delta E_\text{XC}}{\delta n(\vect{r})} = \frac{\partial\epsilon_\text{XC}}{\partial n(\vect{r})} - \vect{\nabla}\cdot\frac{\partial\epsilon_\text{XC}}{\partial\vect{\nabla}n(\vect{r})},
\label{eq:XCPotential}
\end{equation}
where $V^\text{XC}(\vect{r}) \equiv
V^\text{XC}\left[n,\vect{\nabla}n\right](\vect{r})$. Since $V^\text{XC}$ is a
periodic function, we can express its non-equivalent matrix elements in the
real-space basis defined by Eqs.~\eqref{eq:AObasisPeriodic}
and~\eqref{eq:AObasis4c} as the derivative
\begin{equation}
	V^\text{XC}_u = \frac{\partial E_\text{XC}}{\partial D^{\bar{u}}}.
\label{eq:XCMatrix1}
\end{equation}
Applying the chain rule
\begin{equation}
	\frac{\partial E_\text{XC}}{\partial D^{\bar{u}}} = \int_{\reals^3} \frac{\delta E_\text{XC}}{\delta n(\vect{r})} \frac{\partial n(\vect{r})}{\partial D^{\bar{u}}}\dV{r},
\label{eq:XCChainRule}
\end{equation}
and the identity
\begin{equation}
	\Omega_u(\vect{r}) = \frac{\partial n(\vect{r})}{\partial D^{\bar{u}}},
\label{eq:OmegaFromDensity}
\end{equation}
yields
\begin{equation}
	V^\text{XC}_u = \int_{\reals^3} V^\text{XC}(\vect{r})\Omega_u(\vect{r}) \dV{r}.
\label{eq:XCMatrix2}
\end{equation}
Because the integral in Eq.~\eqref{eq:XCMatrix2} is handled numerically, it is
more convenient to use  integration by parts to apply the derivative in the
expression for $V^\text{XC}(\vect{r})$ in Eq.~\eqref{eq:XCPotential} to the
overlap distribution $\Omega_u$. Let us denote
\begin{subequations}
\label{eq:XCNotationDerivativesGroup}
\begin{align}
	V^0_\text{XC}(\vect{r}) &\equiv \frac{\partial\epsilon_\text{XC}}{\partial n(\vect{r})}, & V^i_\text{XC}(\vect{r}) &\equiv \frac{\partial\epsilon_\text{XC}}{\partial\left(\nabla_i n(\vect{r})\right)}, \label{eq:XCNotationDerivatives1}\\
	\Omega_{u,0}(\vect{r}) &\equiv \Omega_{u}(\vect{r}), & \Omega_{u,i}(\vect{r}) &\equiv \nabla_i\Omega_u(\vect{r}), \label{eq:XCNotationDerivatives2}
\end{align}
\end{subequations}
for $i=x,y,z$. Eq.~\eqref{eq:XCMatrix2} can then be written as
\begin{equation}
	V^\text{XC}_u = \int_{\reals^3} V^{\alpha}_\text{XC}(\vect{r})\Omega_{u,\alpha}(\vect{r}) \dV{r},
\label{eq:XCMatrix3}
\end{equation}
where $\alpha=0,x,y,z$. To arrive at a working expression for the XC potential,
we insert the weight functions into Eq.~\eqref{eq:XCMatrix3}, and get
\begin{equation*}
	V^\text{XC}_u = \int_{\reals^3} V_\text{XC}^{\alpha}(\vect{r})\Omega_{u,\alpha}(\vect{r}) \sum_{A\vect{m}'} w_A(\vect{r}-\vect{m}') \dV{r}.
\end{equation*}
It follows that the XC potential becomes
\begin{equation}
	V^\text{XC}_u = \sum_{A\vect{m}'}\int_{\reals^3} V_\text{XC}^{\alpha}(\vect{r})\Omega_{u,\alpha}(\vect{r}+\vect{m}') w_A(\vect{r}) \dV{r}.
\label{eq:XCMatrixFinal}
\end{equation}

\section{\label{sec:ImplementationDetails}Implementation details}

We have implemented the method described in Sec.~\ref{sec:Theory} into the
4c DFT program package \textsc{ReSpect}~\cite{respect5}. Matrix
representations of all operators in real space are obtained by evaluating
the integrals in Eqs.~\eqref{eq:rSpace2cGroup} over the RKB Cartesian GTOs
using the efficient and vectorized integral library
\textsc{InteRest}~\cite{interest}.  The entire implementation is hybrid
OpenMP/MPI parallel, utilizing the OpenMP application programming interface for
intra-node parallelization, and Message Passing Interface (MPI) for inter-node
parallelization.

Before proceeding to the main self-consistent field (SCF) procedure,
\emph{i.e.} the iterative solution of Eq.~\eqref{eq:DKSinMatrixForm}, we
perform these steps:
\begin{itemize}
        \item Exploit the exponential decay of a product of two GTOs
          $\dg{\chi}_{\mu\vect{0}}\chi_{\mu'\vect{m}}$ as their centers become
          more distant in order to generate a \emph{finite} list of significant
          4c overlap distributions.
        \item Form an array of NF unit cells.
        \item Calculate and store the infinite lattice sums $\Lambda_{lm,jk}$
          of the interaction tensor in Eq.~\eqref{eq:latSumOfIntTens} using the
          procedure described in
          Sec.~\ref{sec:TreatmentOfElectrostaticLatticeSums}.
        \item Evaluate the 4c overlap matrix in reciprocal space
          $\tilde{S}(\vect{k})$ in spherical GTOs using
          Eqs.~\eqref{eq:kSpaceOverlap} and~\eqref{eq:rSpaceOverlap4cMat}, and
          orthonormalize the basis applying the L\"{o}wdin canonical
          orthonormalization~\cite{Lowdin-RMP-39-259-1967}, \emph{i.e.} compose
          a transformation matrix $L(\vect{k}) =
          U(\vect{k})\tilde{s}^{-1/2}(\vect{k})$ from the eigenvalues
          $\tilde{s}(\vect{k})$ and eigenvectors $U(\vect{k})$ of
          $\tilde{S}(\vect{k})$. Remove the columns of $L(\vect{k})$ that
          correspond to very small ($<10^{-7}$) eigenvalues
          $\tilde{s}(\vect{k})$ to resolve approximate linear dependencies
          arising in the basis.
\end{itemize}

During the SCF cycle, operators depending on the density matrix must be
reevaluated. The most time-consuming part is the computation of the electron
repulsion integrals (ERIs) of the Coulomb term in
Eq.~\eqref{eq:CoulombMatrixFinalEl} for $\vect{n}$ restricted to the NF unit
cells. Therefore, we employ a variety of approximations and estimates to
accelerate this step. First, centering the multipole expansion at the center of
the overlap distribution $\Omega_u$ that indexes the Fock matrix enables us to
approximate many integrals within the NF using the multipole expansion
\begin{equation}
	J^{\text{e}}_u(\vect{n}) \approx q_u^T(\vect{P})\Theta(\vect{n}-\vect{P})Q,
\label{eq:recenteredMultipoleExp}
\end{equation}
where $\vect{P}$ is the center of $\Omega_u$, and
\begin{equation}
	q_u^{lm}(\vect{P}) = -\int_{\reals^3} \Omega_u(\vect{r})R^{lm}(\vect{r}-\vect{P}) \dV{r},
\label{eq:recenteredMultipoleMomOp}
\end{equation}
is the translated electronic multipole moment operator. Second, we apply the
quaternion adaptation of the Cauchy--Schwarz inequality to obtain an upper
estimate of the remaining ERIs, discarding integrals that contribute negligibly
to the Fock matrix. Details of this integral screening will be published
elsewhere~\cite{Repisky-integralPaper}. Finally, the ERIs that contain a
product of two small-component overlap distributions $\Omega_u^{SS}(\vect{r}_1)
=
\chi^{SS\dagger}_{\mu\vect{0}}(\vect{r}_1)\chi^{SS}_{\mu'\vect{m}}(\vect{r}_1)$
and $\Omega_v^{SS}(\vect{r}_2) =
\chi^{SS\dagger}_{\nu\vect{0}}(\vect{r}_2)\chi^{SS}_{\nu'\vect{n}'}(\vect{r}_2)$
are only computed if: 1) the bra basis function $\mu\vect{0}$ shares the same
center with the ket basis function $\mu'\vect{m}$; and 2) the bra basis
function $\nu\vect{0}$ shares the same center with the ket basis function
$\nu'\vect{n}'$. We denote this scheme as \emph{one-center approximation} to
SS-type ERIs. We tested and tuned these approximations to ensure that the
quality of the results is not affected, and the error introduced by these
approximations is below the error due to the finite basis representation and
numerical integration of the XC term.

To include the XC contributions to the potential and the energy, we calculate
the electronic density
\begin{equation}
	n(\vect{r}) = \sum_{\vect{n}}\Tr\left[ \Omega_u(\vect{r}-\vect{n}) D^{\bar{u}} \right],
\label{eq:densityElXC}
\end{equation}
and its gradients on the DFT grid (see
Sec.~\ref{sec:ExchangeCorrelationContribution}), where the trace is
expressed as in Eq.~\eqref{eq:traceTRsymAD}. The XC potential and its
derivatives $v^{\alpha}(\vect{r})$ are obtained from the \textsc{XCFun}
library~\cite{xcfun} and used to construct the XC Fock matrix elements in
Eq.~\eqref{eq:XCMatrixFinal}.

All relativistic calculations were carried out using a Gaussian finite nucleus
model, as described by Visscher and Dyall~\cite{Visscher-AtData-67-207-1997}.
The finite nucleus model is required in order to regularize the singularity
that appears in the small-component wave function evaluated at the point-type
nuclei; this singularity is otherwise difficult to capture with a finite basis.

The Coulomb and XC contributions are used to assemble the nonzero real-space
quaternion components of the Fock matrix in Eq.~\eqref{eq:rSpaceFock4cMat},
which are then transformed to $\vect{k}$-space, evaluating the Fourier series
in Eq.~\eqref{eq:TRSymOperQuatkSpaceComp}. The 4c $\vect{k}$-space Fock matrix
is composed using Eq.~\eqref{eq:TRSymOperQuat2Cmat}. The kinetic operator is
added in a similar way. The orthonormal basis representation of the Fock matrix
is obtained as $F(\vect{k})\rightarrow\dg{L}(\vect{k})F(\vect{k})L(\vect{k})$.
The Fock matrix is diagonalized, and from its band energies
$\epsilon_p(\vect{k})$, an occupation vector $f_p(\vect{k})$ is formed
[Eq.~\eqref{eq:FermiDirac}]. The $\vect{k}$-space density matrix is obtained in
the orthonormal basis according to Eq.~\eqref{eq:kSpaceDmat2}, and transformed
as $D(\vect{k})\rightarrow L(\vect{k})D(\vect{k})\dg{L}(\vect{k})$.

The new density matrix in real space $D^{\mu\vect{m},\mu'\vect{0}}$ is
constructed by calculating the integral in Eq.~\eqref{eq:rSpaceDmatQuadrature}
over the first Brillouin zone. The integral is approximated by a sum over a
$\Gamma$-centered uniform mesh of $\vect{k}$ points with equal weights
$|\mathcal{K}|/\mathcal{N}$, where $\mathcal{N}$ is the total number of sampled
$\vect{k}$ points. Specifically, let $\vect{b}_i$ denote the primitive vectors
in reciprocal space for $i=1,\ldots,d$. Then the mesh consists of
$\vect{k}$ points defined as
\begin{align}
	\vect{k} &= \sum_{i=1}^d \frac{k_i}{\mathcal{N}_i} \vect{b}_i, & k_i &=-\frac{\mathcal{N}_i-1}{2},\ldots,\frac{\mathcal{N}_i-1}{2}, \label{eq:kSpaceGrid}
\end{align}
where $\mathcal{N}_i$ is the total number of $\vect{k}$ points in the $i$-th
crystallographic direction. Such an integration scheme does not capture the
discontinuity of the integrand at the Fermi surface arising in metallic
systems. However, in this work we study systems with a nonzero band gap, and
the integration scheme proved sufficiently accurate.

In order to accelerate the SCF convergence, we extrapolate the real-space Fock
matrix using the linear combination of Fock matrices from the current and the
previous SCF cycles, before transforming it to reciprocal space. The
extrapolation coefficients are determined from the direct inversion of the
iterative subspace (DIIS) procedure of
Pulay~\cite{Pulay-CPL-73-393-1980,Pulay-JCC-3-556-1982}, applied only to the
$\Gamma$-point ($\vect{k}=\vect{0}$), \emph{i.e.} using error vectors defined
as $e = \left[ F(\vect{0}), D(\vect{0})\right]$ (in the orthonormal basis).
Such a restriction has been demonstrated to be satisfactory for solid-state
calculations~\cite{Kudin-phdthesis-2000,Lazarski-JCTC-11-3029-2015,Maschio-TCA-137-60-2018}.

\section{\label{sec:ProblemsWithDiffuseFunctions}Problems with diffuse functions}

\subsection{\label{sec:NonrelativisticTheory}Nonrelativistic theory}

Gaussian basis functions with diffuse exponents are known to cause numerical
instabilities in the SCF procedure for
solids~\cite{Suhai-CP-68-467-1982,Geipel-CPL-273-62-1997,Kudin-PRB-61-16440-2000,Jacquemin-IJQC-89-452-2002,Peintinger-JCC-34-451-2013}.
One type of instability is associated with the overcompleteness of a chosen
basis, \emph{i.e.} ``true'' linear dependence of the basis that usually occurs
when the smallest eigenvalue of the overlap matrix is below a certain threshold
($10^{-7}$)~\cite{Suhai-CP-68-467-1982,Jacquemin-IJQC-89-452-2002,Kudin-PRB-61-16440-2000}.
We remove such linear dependencies during the basis orthonormalization step by
applying the procedure described in Sec.~\ref{sec:ImplementationDetails}.
Another type of instability arises when the Fock matrix elements are calculated
with large errors, \emph{e.g.} due to a premature truncation of the infinite
lattice sums (see Sec.~\ref{sec:TreatmentOfElectrostaticLatticeSums}). This
problem was reported if the lowest eigenvalue of the overlap matrix was below
$10^{-2}$~\cite{Suhai-CP-68-467-1982,Jacquemin-IJQC-89-452-2002,Kudin-PRB-61-16440-2000}.
For these reasons, it is a common practice to exclude most diffuse functions
from solid-state calculations altogether, either by deleting them from the
molecular basis
sets~\cite{Geipel-CPL-273-62-1997,Peralta-JCP-122-084108-2005,Peintinger-JCC-34-451-2013,Zhao-JCP-144-044105-2016}
(a rule of thumb is to remove exponents smaller than 0.1) or by reoptimizing
the basis set exponents and the contraction
coefficients~\cite{Peintinger-JCC-34-451-2013}.

In this work, we did not encounter the aforementioned problems in the
nonrelativistic implementation, and deleting the diffuse functions from the
basis set proved to be unnecessary.  On the contrary, we observed that removing
the diffuse functions produced significant errors in some of the calculated
band gaps (see Table~\ref{tab:AgXresultsGapsBasis}), and the systematic
convergence of so-constructed basis sets was lost. Thus we recommend caution
when making such severe modifications of basis sets.  We believe that the
problems with diffuse functions can be mitigated by proper handling of the
lattice sums.

\subsection{\label{sec:RelativisticTheory}Relativistic theory}

While the nonrelativistic implementation did not pose convergence challenges
even with the original unmodified molecular basis with diffuse functions, the
same is not true in the 4c case.  For the three-dimensional silver halides
examined here, we found that the energy gap between the negative- and
positive-energy states was closed. In fact, a small number of the
negative-energy states was located in the energy region of the occupied
electronic states. Occupying any of these intruder states disrupted the SCF
procedure and made it impossible to reach convergence. This pathological behavior
occurred even if the magnitude of the lowest eigenvalues of the small-component
overlap (kinetic) matrix $\mathcal{T}$ was of the order of $10^{-4}$, and the
behavior was not observed if the diffuse functions were excluded from the
calculation.

To understand this problem, let us study a model Dirac Hamiltonian expressed in
an RKB basis containing one basis function. In absence of SOC,
Eq.~\eqref{eq:DKSinMatrixForm} can be written as the $2\times 2$ equation
\begin{equation}
	\begin{pmatrix}
		v & \frac{1}{2}t \\
		\frac{1}{2}t & \frac{w}{4c^2}-\frac{1}{2}t
	\end{pmatrix} \begin{pmatrix}
		c_L \\ c_S
	\end{pmatrix} = \epsilon \begin{pmatrix}
		s & 0 \\
		0 & \frac{t}{4c^2}
	\end{pmatrix} \begin{pmatrix}
		c_L \\ c_S
	\end{pmatrix},
\label{eq:pos:DiracEqModel2x2}
\end{equation}
where $s, t\in\reals^{+}$ parametrize the 4c overlap matrix in
Eq.~\eqref{eq:rSpaceOverlap4cMat}; $v,w\in\reals$ are the large and
small-component contributions to the potential in
Eq.~\eqref{eq:rSpaceFock4cMat}, respectively; $\epsilon$ is an eigenvalue; and
$\left(c_L, c_S\right)^T$ is an eigenvector. Here, we omit the dependence on
$\vect{k}$, as it is not relevant for the following discussion. The
orthonormalized Hamiltonian thus becomes
\begin{equation}
	H = \begin{pmatrix}
		\frac{v}{s} & \sqrt{\frac{t}{s}}c \\
		\sqrt{\frac{t}{s}}c & \frac{w}{t}-2c^2
	\end{pmatrix}.
\label{eq:pos:DiracHam2x2}
\end{equation}
Expanding the solutions of this Hamiltonian as $c\rightarrow\infty$ gives
\begin{subequations}
\begin{align}
	\epsilon_{+}(c) &= \frac{v}{s} + \frac{t}{2s} + O\left(\frac{1}{c^2}\right), \label{eq:pos:solutions2x2cPositive}\\
	\epsilon_{-}(c) &= -2c^2 + \frac{w}{t} - \frac{t}{2s} + O\left(\frac{1}{c^2}\right). \label{eq:pos:solutions2x2cNegative}
\end{align}
\label{eq:pos:solutions2x2c}
\end{subequations}
Similarly, the asymptotic expansion of the solutions as $t\rightarrow 0$ gives
\begin{subequations}
\begin{align}
	\epsilon_{+}(t) &= \frac{v}{s} + O\left(t^2\right), \label{eq:pos:solutions2x2tPositive}\\
	\epsilon_{-}(t) &= -2c^2 + \frac{w}{t} + O\left(t^2\right). \label{eq:pos:solutions2x2tNegative}
\end{align}
\label{eq:pos:solutions2x2t}
\end{subequations}
The expansion in Eq.~\eqref{eq:pos:solutions2x2cNegative} shows that
$\epsilon_{-}$ is singular as $c\rightarrow\infty$, whereas according to
Eq.~\eqref{eq:pos:solutions2x2tNegative}, $\epsilon_{-}$ is also singular as
$t\rightarrow 0$. This is in contrast with $\epsilon_{+}$ which does not
exhibit such singularities. If $w>0$ then the term $\frac{w}{t}$ increases the
energy of $\epsilon_{-}$. This increase can become significant for large values
of $w$ or, equivalently, small values of $t$, and can shift the negative-energy
state to the electron region close to $\epsilon_{+}$. In practical
calculations, the Coulomb potential consists of both the electron--nuclear
attraction as well as the electron--electron repulsion. While the attractive
Coulomb potential gives rise to bound states just below the positive-energy
continuum, the repulsive potential produces bound states just above the
negative-energy continuum~\cite{Ayenew-msthesis-2000}. The matrix $V$ in
Eq.~\eqref{eq:rSpaceTmatAndVmat4comp} is indefinite, \emph{i.e.} with both
negative and positive eigenvalues, and the behavior corresponding to $w>0$ can
be observed. Some of the highest-lying (spurious) negative-energy states can
thus have a higher energy than the lowest positive-energy states. This
``inverse variational collapse'' is possible because the RKB basis only
guarantees that the low-lying positive-energy bound states do not collapse into
the negative-energy continuum (except for superheavy elements with large values
of $Z$~\cite{Schwerdtfeger-NPA-944-551-2015}), but does not prevent the
negative-energy bound states from intruding the positive-energy
region~\cite{Quiney-PS-36-460-1987,Shabaev-PRL-93-130405-2004,Sun-TCA-129-423-2011}.

Since the negative- and positive-energy states can overlap if diffuse functions
are included in basis sets used for solid-state calculations, the conventional
procedure of forming the occupation vector in Eq.~\eqref{eq:FermiDirac} by
assuming that the electronic bound states are well-separated from the
negative-energy states~\cite{Saue-CPC-12-3077-2011} is not justified.
Expansions in Eqs.~\eqref{eq:pos:solutions2x2c} indicate that it is possible to
identify the negative-energy states by probing their dependence on the speed of
light. Here, we perturb the one-electron Dirac Hamiltonian by infinitesimally
shifting the square of the speed of light, \emph{i.e.} we employ the
substitution $c^2\rightarrow c^2(1+\lambda)$ in the Fock operator expressed in
the orthonormalized basis, and evaluate $\xi_p(\vect{k}) \equiv
\left.\frac{1}{2c^2}\pd{\epsilon_p(\vect{k},\lambda)}{\lambda}\right|_{\lambda=0}$.
We found that the states with negative values of $\xi_p(\vect{k})$ must be left
vacant in order to converge the SCF procedure. The negative-energy states that
penetrated into the positive-energy spectrum always appeared in pairs: One
virtual state with a high energy and $\xi_p(\vect{k})\approx -1$, and one
orbital in the region of occupied electron states with $-1<\xi_p(\vect{k})<0$,
presumably corresponding to a bound negative-energy state. These intruder
states did not appear in calculations on finite systems consisting of one unit
cell (molecule). A more robust approach to mitigate this problem will be a
subject of further research.

\section{\label{sec:Results}Results}

To asses the performance of the proposed methodology, we have performed energy
band-gap calculations at different $\vect{k}$ points for the three-dimensional
silver halides (Ag$X$, $X$=Cl, Br, I) using both fully relativistic (4c) and the
nonrelativistic one-component (1c) density functional level of theory. Despite of
their highly symmetric cubic fcc structure, Ag$X$ serve as an excellent probe for
the 4c method for a number of reasons. The unit cell of Ag$X$ has a nonzero
dipole moment, and the Coulomb lattice sums exhibit the most complicated,
conditional convergence. In addition, silver halides are small-gap indirect
semiconductors~\cite{Peralta-JCP-122-084108-2005,Zhao-JCP-144-044105-2016} with
a densely packed structure, and thus pose more challenges to the SCF procedure
as well as to the employed basis sets. Finally, fully relativistic 4c
calculations using simulation supercells that contain more than six hundred
heavy atoms and tens of thousand electrons are memory and CPU demanding.

Even though the ionic Ag$X$ crystals exhibit large relativistic effects, these
are predominantly of a scalar-relativistic origin while SOC plays only a minor
role~\cite{Peralta-JCP-122-084108-2005,Zhao-JCP-144-044105-2016}.  To better
assess how well our approach can treat SOC effects, we also study the
two-dimensional graphene-like honeycomb structures of silicene and
germanene~\cite{Liu-PRL-107-076802-2011} that possess a large SOC-driven
quantum spin Hall effect.

\subsection{\label{sec:SilverHalides}Silver halide crystals}

Equilibrium lattice constants of Ag$X$ were taken from the recent work of Zhao
{\it et al.\/}~\cite{Zhao-JCP-144-044105-2016}, and the nonrelativistic
GGA-type XC functional PBE~\cite{Perdew-PRL-77-3865-1996} was employed.
The numerical integration of the XC contributions was performed on a grid
consisting of 302 angular points for each atom, 80 radial points for Ag, and 70
radial points for the halides. Reciprocal space integration was evaluated on a
uniform mesh of $7\times 7\times 7$ $\vect{k}$ points
[Eq.~\eqref{eq:kSpaceGrid}]. For the large-component basis, the all-electron
pob-TZVP basis set of triple-$\zeta$ quality optimized for solid-state
calculations~\cite{Peintinger-JCC-34-451-2013} was used; however, the basis was
uncontracted, as is required for relativistic calculations, we denote this
basis as upob-TZVP. Since upob-TZVP is not available for heavier elements, we
employed the uncontracted all-electron double-$\zeta$ (DZ) basis sets of
Dyall~\cite{Dyall-TCA-99-366-1998,*[addendum
]Dyall-TCA-108-365-2002-ad,*Dyall-TCA-108-335-2002,*[revision
]Dyall-TCA-115-441-2006,Dyall-TCA-117-483-2007} for Ag and I. The
small-component basis functions were generated on-the-fly using the RKB
condition in Eq.~\eqref{eq:AObasis4cSS}.

\begin{table}[tb]
\caption{\label{tab:AgXresultsGaps} Energy band gaps of three-dimensional Ag$X$
  systems obtained for various $\vect{k}$ points at the fully relativistic (fr)
  and nonrelativistic (nr) level of theory using the PBE XC functional.  The
  upob-TZVP basis was employed for Cl and Br, and Dyall's double-$\zeta$ for Ag
  and I, both with (DZ) and without (rDZ) the most diffuse functions.}
\begin{ruledtabular}
\begin{tabular}{ccccccc}
      &             &       & \multicolumn{4}{c}{Gap [eV]} \\\cline{4-7}
 AgCl & $a_0$ [\AA] & Basis & L--L & $\Gamma$--$\Gamma$ & X--X & L--$\Gamma$\\
\hline
nr & 5.692 & rDZ                  & 5.18 & 3.53 & 5.45 & 1.74 \\
   &       &  DZ                  & 4.93 & 3.47 & 5.47 & 1.68 \\
   &       & STO\footnotemark[1]  & 4.72 & 3.44 & 5.29 & 1.67 \\
   &       & LAPW\footnotemark[2] & 4.76 & 3.44 & 5.29 & 1.69 \\
fr & 5.612 & rDZ                  & 4.67 & 2.95 & 4.20 & 0.89 \\
   &       &  DZ                  & 4.47 & 2.93 & 4.20 & 0.87 \\
   &       & STO\footnotemark[1]  & 4.27 & 2.99 & 4.03 & 0.88 \\
   &       & LAPW\footnotemark[2] & 4.30 & 3.02 & 4.04 & 0.89 \vspace{0.1cm}\\
 AgBr & $a_0$ [\AA] & Basis & L--L & $\Gamma$--$\Gamma$ & X--X & L--$\Gamma$\\
\hline
nr & 5.937 & rDZ                  & 4.76 & 3.15 & 4.83 & 1.77 \\
   &       &  DZ                  & 4.36 & 2.96 & 4.81 & 1.59 \\
   &       & STO\footnotemark[1]  & 4.31 & 2.97 & 4.81 & 1.57 \\
   &       & LAPW\footnotemark[2] & 4.35 & 2.96 & 4.79 & 1.58 \\
fr & 5.843 & rDZ                  & 4.13 & 2.34 & 3.67 & 0.70 \\
   &       &  DZ                  & 3.82 & 2.24 & 3.68 & 0.61 \\
   &       & STO\footnotemark[1]  & 3.77 & 2.25 & 3.67 & 0.60 \\
   &       & LAPW\footnotemark[2] & 3.82 & 2.24 & 3.68 & 0.61 \vspace{0.1cm}\\
 AgI  & $a_0$ [\AA] & Basis & L--L & $\Gamma$--$\Gamma$ & X--X & L--X\\
\hline
nr & 6.280 & rDZ                  & 5.12 & 3.28 & 3.58 & 1.62 \\
   &       &  DZ                  & 3.99 & 3.11 & 3.54 & 1.59 \\
   &       & STO\footnotemark[1]  & 3.91 & 3.14 & 3.56 & 1.60 \\
   &       & LAPW\footnotemark[2] & 3.92 & 3.13 & 3.54 & 1.58 \\
fr & 6.169 & rDZ                  & 4.14 & 1.96 & 2.75 & 0.50 \\
   &       &  DZ                  & 3.25 & 1.88 & 2.74 & 0.49 \\
   &       & STO\footnotemark[1]  & 3.17 & 1.90 & 2.76 & 0.49 \\
   &       & LAPW\footnotemark[2] & 3.18 & 1.91 & 2.75 & 0.47 \\
\end{tabular}
\end{ruledtabular}
\footnotetext[1]{Ref.~\cite{Zhao-JCP-144-044105-2016}, 2c X2C approach.}
\footnotetext[2]{Ref.~\cite{Zhao-JCP-144-044105-2016}, 4c approach.}
\end{table}

\begin{figure*}[tb]
\includegraphics[width=1.00\textwidth]{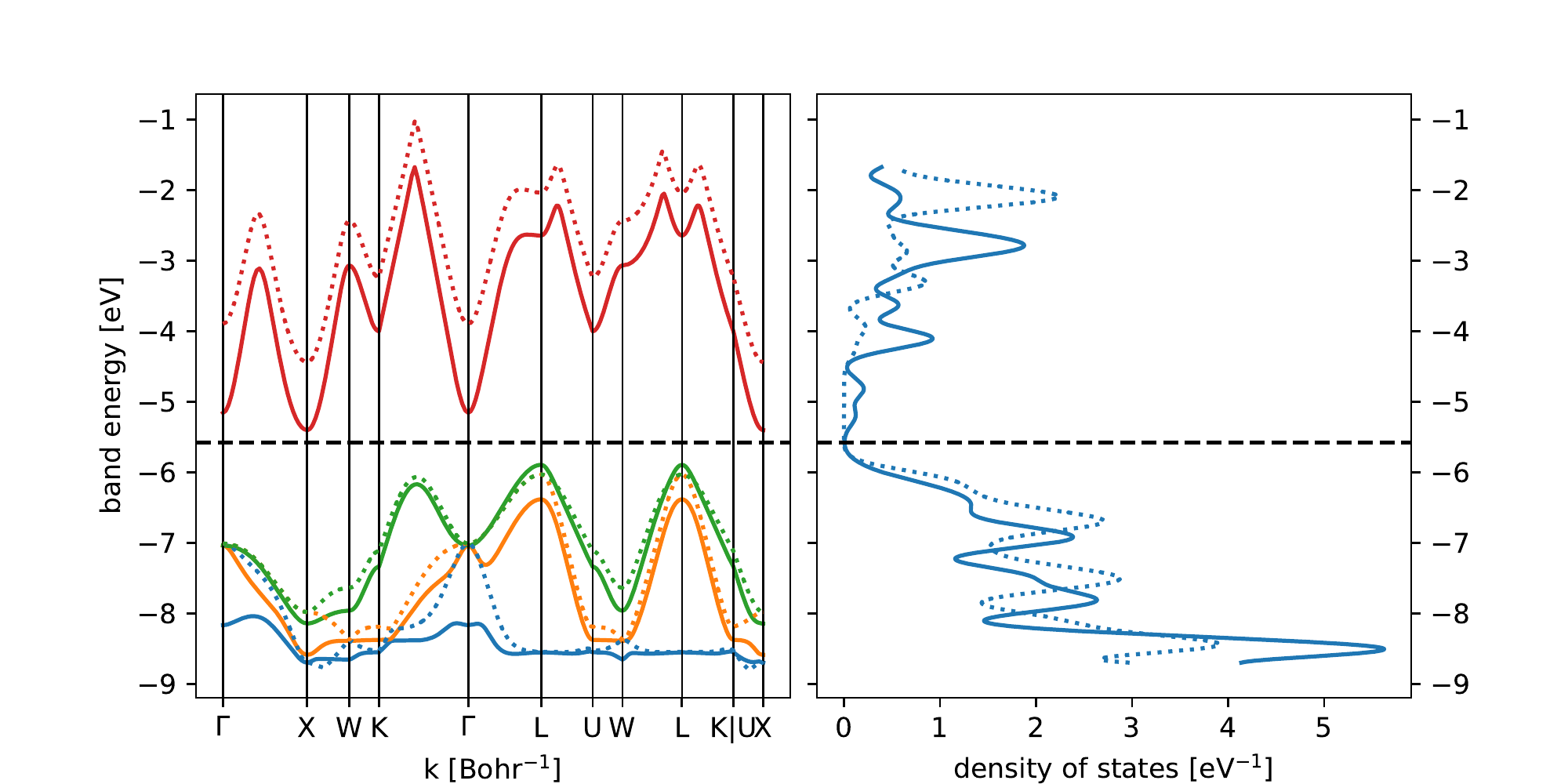}
\caption{\label{fig:AgIBands} Band structure diagram (left) and the DOS (right)
  of AgI obtained at the 4c (full line) and 1c (dashed line) levels of theory
  with the PBE XC functional. The horizontal dashed black line separates the
  occupied and the vacant states, and was placed in the middle of the band gap.
  The path traversing high-symmetry $\vect{k}$-points in the reciprocal-space
  unit cell was chosen according to Ref.~\cite{Setyawan-CMS-49-299-2010}. The
  figure was generated using Python matplotlib library~\cite{matplotlib}.}
\end{figure*}

In order to alleviate the convergence problems described in
Sec.~\ref{sec:RelativisticTheory} that are related to diffuse Gaussian-type
functions in the atomic basis sets, we followed the common practice of removing
the GTOs with exponents $<0.1$ from the original basis
sets~\cite{Geipel-CPL-273-62-1997,Peralta-JCP-122-084108-2005,Peintinger-JCC-34-451-2013,Zhao-JCP-144-044105-2016}.
In this work, we deleted the most diffuse s- and p-type functions on Ag, and
denoted the reduced basis by acronym ``r'' added in front of the original basis
set name. During the numerical integration of the XC term, GTOs were considered
to have a finite extent, defined as the radius of an atom-centered sphere
outside of which values of the most diffuse Gaussian function are below a
user-defined threshold. The extent of the original basis sets was 12.1~\AA{}
due to the diffuse functions on Ag. Using the truncated r-type basis sets
reduced this extent to values between 7.4 and 7.8~\AA{}. The GTO extent
defined as in Ref.~\cite{Watson-JCP-121-2915-2004} used in the multipole
expansions of the Coulomb term had almost identical values.

Table~\ref{tab:AgXresultsGaps} shows the results of our 4c and 1c calculations
of the energy gaps for the Ag$X$ systems. Our values are compared with the
results calculated using two different
techniques~\cite{Zhao-JCP-144-044105-2016}: 2c method based on the X2C
Hamiltonian and STOs, and the 4c LAPW method.  The vertical (direct) band gaps
are obtained at a set of special $\vect{k}$-points: $\Gamma$, L, and X.  The
band-structure diagram and the density of states (DOS) of AgI calculated at the
1c and 4c levels are depicted in Figure~\ref{fig:AgIBands}; DOS was obtained as
$N(\epsilon) \equiv \frac{1}{\mathcal{N}}\sum_{p\vect{k}}
\delta(\epsilon-\epsilon_{p\vect{k}})$, where $\mathcal{N}$ is the total number
of sampled $\vect{k}$ points, and the $\delta$-function was represented with a
Gaussian with a standard deviation of 136 meV.  These results show that the
ionic Ag$X$ compounds are indirect semi-conductors, with the band gap occurring
between the L and $\Gamma$ points for AgCl and AgBr, and between the L and X
points for AgI. This agrees with the findings of previous
studies~\cite{Peralta-JCP-122-084108-2005,Zhao-JCP-144-044105-2016}.  All band
gaps are significantly reduced when including relativistic effects, and
Figure~\ref{fig:AgIBands} reveals that this reduction is due to a large
decrease in the energy of the entire conduction band. In
Figure~\ref{fig:AgIBands}, we can also observe strong SOC splittings that occur
within the valence bands. In particular, SOC lifts the degeneracy on the
$\Gamma$-X and $\Gamma$-L lines, and for $\vect{k}=\Gamma$,  the difference
between the split energies equals to 1.13 eV. Overall, our results calculated
with the DZ basis set agree well with those presented in
Ref.~\cite{Zhao-JCP-144-044105-2016}; we reproduce the general trends as well
as the difference between the relativistic and the nonrelativistic
calculations.

\begin{table}[tb]
\caption{\label{tab:AgXresultsGapsBasis} Nonrelativistic 1c energy band gaps of
  three-dimensional AgI calculated with a hierarchy of basis sets. All results
  are obtained with the PBE functional.  Comparisons are made with
  non-relativistic literature values obtained using either STOs or plane waves,
  and are taken from Ref.~\cite{Zhao-JCP-144-044105-2016}.}
\begin{ruledtabular}
\begin{tabular}{ccccc}
AgI    & \multicolumn{4}{c}{Gap [eV]} \\\cline{2-5}
basis  & L--L & $\Gamma$--$\Gamma$ & X--X & L--X\\
\hline
rDZ                  & 5.12 & 3.28 & 3.58 & 1.62 \\  
rVDZ                 & 4.87 & 3.30 & 3.60 & 1.62 \\  
rVTZ                 & 4.66 & 3.74 & 3.50 & 1.56 \\  
rVQZ                 & 4.01 & 3.21 & 3.56 & 1.59 \\  
 DZ                  & 3.99 & 3.11 & 3.54 & 1.59 \\  
 VDZ                 & 3.95 & 3.14 & 3.56 & 1.59 \\  
STO\footnotemark[1]  & 3.91 & 3.14 & 3.56 & 1.60 \\  
LAPW\footnotemark[1] & 3.92 & 3.13 & 3.54 & 1.58 \\  
\end{tabular}
\end{ruledtabular}
\footnotetext[1]{Ref.~\cite{Zhao-JCP-144-044105-2016}.}
\end{table}

On the other hand, there is a notable discrepancy between the L--L direct gaps
evaluated using the DZ and the rDZ basis sets, particularly for AgI. The fact
that DZ results agree well with the STO and LAPW results in
Ref.~\cite{Zhao-JCP-144-044105-2016} indicates that the diffuse functions are
of immense importance for the band structures of these systems, and should not
be removed from the basis set. To further investigate the basis set effect, we
conducted additional tests at the 1c level with various basis sets, and the
results for band gaps of AgI are summarized in
Table~\ref{tab:AgXresultsGapsBasis}. In addition to the DZ basis set, we
included the larger Dyall's valence double-$\zeta$ (VDZ), as well as the
hierarchical system of basis sets: reduced valence double-, triple-, and
quadruple-$\zeta$ (rVDZ, rVTZ, rVQZ), with discarded the exponents smaller than
$0.1$. The sequence of basis sets without the diffuse functions does not
exhibit an apparent convergence, the $\Gamma$--$\Gamma$ gap deviates more from
the reference results for larger basis sets. Acceptable agreement is reached
only with the very large rVQZ. This issue does not appear for the original
basis sets with diffuse exponents, and our results agree very well with those
of Zhao \emph{et al.}~\cite{Zhao-JCP-144-044105-2016} already for DZ and VDZ.
A similar observation was done by Zhao \emph{et al.}, who performed test
calculations on AgCl with polarized double-$\zeta$ STOs, and the calculated
band gaps differed marginally ($<0.1$ eV) from the results obtained with the
large reduced polarized quadruple-$\zeta$ basis (rQZ4P) with eliminated diffuse
s and p functions.  In addition, this is in line with the findings of Te Velde
and Baerends~\cite{TeVelde-PRB-44-7888-1991} that a reasonable basis-set limit
(with errors $<10^{-3}$ a.u. in cohesive energies per atom) can  be reached for
densely packed systems already with STOs of double-$\zeta$ quality, provided
they contain polarization functions.  Considering that GTOs and STOs only
differ in the radial part, one would expect that a similar behavior should be
seen also for GTOs.  We have confirmed this observation, but only for the full
original DZ basis set with diffuse exponents.  Therefore, great care must be
taken when adopting basis sets for solid-state calculations, and we do not
generally recommend deleting diffuse exponents for heavy elements. Optimized
solid-state GTOs have been developed by Peintinger \emph{et
al.}~\cite{Peintinger-JCC-34-451-2013} for the lighter elements of the periodic
table, but this work needs to be extended to address the elements in the lower
part of the periodic table as well.

\begin{figure}[tb]
\includegraphics[width=1.00\columnwidth]{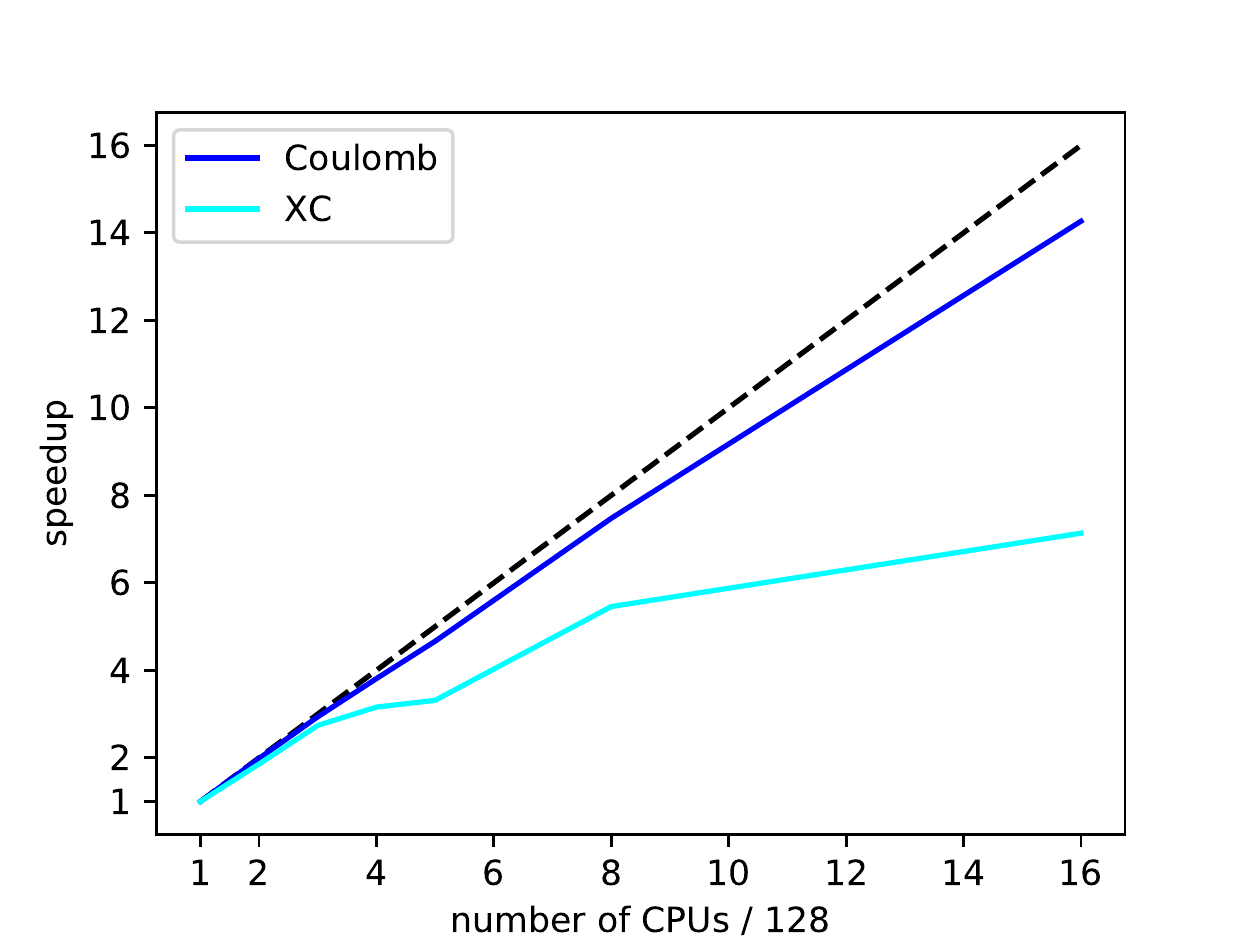}
\caption{\label{fig:AgIScaling} Speedup on the wall-clock time of the
  evaluation of the Coulomb and XC contributions to the Fock matrix in real
  space for 1 SCF cycle of the 4c AgI as a function of the number of CPUs used.
  The reference calculation was performed using 128 CPUs (4 nodes), and the
  largest calculation used 2048 CPUs (64 nodes).  The dashed line denotes a
  hypothetical (linear) scaling of wall-clock time with given computational
  resources. The figure was generated using Python matplotlib
  library~\cite{matplotlib}.}
\end{figure}

Finally, we tested the parallel performance of our implementation with respect
to the number of central processing units (CPUs) used. We conducted a series of
4c calculations on AgI with the DZ basis using two-socket computational nodes,
where each socket consists of 16 physical cores. We used the calculation on
four nodes (128 CPUs) as a reference.  Figure~\ref{fig:AgIScaling} demonstrates
near-ideal linear scaling with the number of processors of the NF Coulomb
contributions to the Fock matrix and energy. The implementation remains
efficient even when 64 nodes (2048 CPUs) are used. The evaluation of the XC
contributions exhibits optimal scaling for a smaller number of nodes, but
becomes less optimal beyond 1024 CPUs. However, absolute wall-clock times
required to compute the XC contribution are considerably shorter than for the
Coulomb terms. The remaining steps in the algorithm, such as the FF Coulomb
contributions, Fourier transformation, matrix diagonalizations, and the DIIS,
are negligible.

\subsection{\label{sec:HoneycombStructures}Honeycomb structures}

To validate our method on systems displaying larger spin--orbit effects, we
have also calculated the band structure of the heavier two-dimensional analogues
of graphene: silicene and germanene. Both systems have been found to be stable
in a low-buckled hexagonal geometry~\cite{Liu-PRL-107-076802-2011}, contrary to
the truly planar graphene. In contrast to graphene, the buckled geometry of
silicene and germanene enhances the SOC effect~\cite{Liu-PRL-107-076802-2011}.
To compare our calculated  band gaps with literature values, we used the
geometries from Ref.~\cite{Liu-PRL-107-076802-2011}, and the nonrelativistic
PBE functional~\cite{Perdew-PRL-77-3865-1996}. The integration grid for the XC
contributions contained 80 radial points per atom, and Lebedev quadrature grid
points of an adaptive size in the angular part~\cite{Krack-JCP-108-3226-1998}.
Reciprocal space integration was performed on a uniform grid of $31\times 31$
$\vect{k}$-points. We studied the effect of basis set on the band gap, and
employed the uncontracted all-electron
upob-TZVP~\cite{Peintinger-JCC-34-451-2013} and the hierarchy of systematically
improved Dunning's basis sets~\cite{Wilson-JCP-110-7667-1999} (ucc-pVDZ,
ucc-pVTZ, ucc-pVQZ).

\begin{table}[tb]
\caption{\label{tab:2DresultsGaps}
Band gaps of two-dimensional honeycomb structures at the fully relativistic (fr)
  4c and nonrelativistic (nr) 1c level of theory using the PBE functional and
  various basis sets.  Geometries are taken from
  Ref.~\cite{Liu-PRL-107-076802-2011}.}
\begin{ruledtabular}
\begin{tabular}{ccdd}
 & & \multicolumn{2}{c}{\textrm{Band gap [meV]}} \\\cline{3-4}
 Method & Basis & \multicolumn{1}{c}{\textrm{Silicene}} & \multicolumn{1}{c}{\textrm{Germanene}} \\
\hline
nr & upob-TZVP & 0.026 &  0.028\\
fr & upob-TZVP & 1.548 & 25.119\\
fr &  ucc-pVDZ & 1.596 & 24.296\\
fr &  ucc-pVTZ & 1.606 & 24.323\\
fr &  ucc-pVQZ & 1.607 & 24.342\\
Ref.~\cite{Liu-PRL-107-076802-2011} && 1.55 & 23.9 \\
\end{tabular}
\end{ruledtabular}
\end{table}

Table~\ref{tab:2DresultsGaps} collects our calculated band gaps at the 1c and
4c levels of theory at the Dirac points ($\vect{k}=K$) of silicene and
germanene. For comparison, we report in Table~\ref{tab:2DresultsGaps} also the
results of Liu, Feng and Yao~\cite{Liu-PRL-107-076802-2011} calculated using
the relativistic pseudopotential PAW
approach~\cite{DalCorso-PRB-82-075116-2010}.  Since these graphene-like
structures exhibit a quantum spin Hall
effect~\cite{Gmitra-PRB-80-235431-2009,Liu-PRL-107-076802-2011,Han-NatureNano-9-794-2014},
the existence of a nonzero gap is solely due to SOC. Hence, the nonrelativistic
band gaps should then be strictly zero. The numbers in
Table~\ref{tab:2DresultsGaps} do not display this feature exactly, but we
attribute the very small values of the nonrelativistic gaps to numerical noise
and the truncation of the expansion of the one-electron bases (finite basis
effect). The convergence with respect to the basis limit is very fast --- our
ucc-pVDZ band gaps differ only marginally from the ucc-pVTZ and ucc-pVQZ
results, whereas a larger discrepancy exists already between the ucc-pVDZ
results and the results in Ref.~\cite{Liu-PRL-107-076802-2011}. Since our band
gaps are obtained at the fully relativistic 4c level with the all-electron
potential and well-converged basis, one can consider them as reference data.
The small discrepancy between our 4c method and the previously reported 2c
Pauli-type relativistic PAW method~\cite{DalCorso-PRB-82-075116-2010} of
Ref.~\cite{Liu-PRL-107-076802-2011} can be attributed to the different
treatment of relativity and the use of the pseudopotential approximation in the
latter approach.  We believe that these results demonstrate that the presented
methodology opens a new possibility to study heavy-element-containing materials
with promising technological applications, for instance in spintronic devices.

\section{\label{sec:Conclusion}Conclusion and outlook}

We have presented a first-principles full-potential relativistic method and its
implementation for solving the 4c Dirac--Kohn--Sham equation for periodic
systems employing a local basis composed of Gaussian-type orbitals.  The
proposed method accounts variationally for both scalar-relativistic as well as
spin--orbit effects,  allowing us to study solids across the entire periodic
table in a uniform and consistent manner. The explicit built-in periodicity
allows for a treatment of systems of arbitrary dimensionality without having to
introduce nonphysical replicas of the systems studied in non-periodic
dimensions. We formulated key principles of the method in the 4c
Kramers-restricted framework, exploiting the time-reversal structure of
operators in real and reciprocal space, and showed how to assemble the
real-space Coulomb and exchange--correlation operators in this framework.  We
have discussed the conditionally-convergent electrostatic infinite lattice sums
arising in studies of periodic systems, and we adopted the multipole expansion
and an iterative renormalization procedure to calculate the lattice sums of the
interaction tensor. To accelerate the calculations, some explicit two-electron
integrals were neglected based on an efficient screening scheme, or
approximated with  a multipole expansion. We have analyzed the problem of
inverse variational collapse that emerges in the 4c method if the employed
basis set contains diffuse functions, and have suggested a means for avoiding
the breakdown of the 4c SCF procedure.  The method has been implemented in the
4c \textsc{ReSpect}~\cite{respect5} code, using the vectorized integral library
\textsc{InteRest}~\cite{interest}. Finally, we have validated this methodology
on some exemplary calculations of three-dimensional silver halide crystals in their
fcc phase, and two-dimensional honeycomb structures featuring the quantum spin
Hall effect.  Energy band gaps were calculated at various special
$\vect{k}$-points. Overall, our results agreed very well with earlier published
findings. Furthermore, we have demonstrated that the convergence with respect
to the basis limit is possible for standard basis sets used for molecular
calculations in quantum chemistry, without the need to modify the basis sets by
removing the most diffuse exponents. We obtained very good cost--performance
ratio of our hybrid OpenMP/MPI parallel implementation as we increased the
number of used CPUs up to 2048.

The methodology presented in this paper holds promise in the computational
study of solid-state materials.  The 4c scheme is conceptually simpler and more
transparent than approximate 2c techniques, and can be used to produce
reference results to benchmark more approximate methods, and in this way
increase confidence in approximate schemes and thus pave the way for
computational studies of more complex materials. Furthermore, the
full-potential formalism adopted here enables investigations of unique features
of spin--orbit coupled materials, such as magnetic response properties and
core-electron (X-ray) spectroscopy, where a full relativistic description is
needed. We also believe that the method can prove valuable in a search for
materials with non-trivial topological properties.

\begin{acknowledgments}
This work was supported by the Research Council of Norway through its Centres
  of Excellence scheme (Grant Nos. 179568 and 262695), as well as through a
  research grant (Grant No. 214095). Support from the Troms\o\ Research
  Foundation is also gratefully acknowledged. Computer time was provided by the
  Norwegian Supercomputer Program NOTUR (Grant No. NN4654K) as well as by the
  Large Infrastructures for Research, Experimental Development and Innovations
  project ``IT4Innovations National Supercomputing Center -- LM2015070''
  (Project No. OPEN-12-40) supported by The Ministry of Education, Youth and
  Sports of the Czech Republic. The publication charges for this article have
  been funded by a grant from the publication fund of UiT The Arctic University
  of Norway. MK would like to acknowledge Stanislav Komorovsky, Stefan Varga
  and Marc Joosten for fruitful discussions on the topic.
\end{acknowledgments}

\appendix
\section{\label{app:TranslationSymmetry}Translational symmetry}

In this appendix, we review some consequences of the translational symmetry on
operators in various basis representations. We will here only be concerned with
discrete translations, \emph{i.e.} translations by an arbitrary
integer-modulated lattice vector $\vect{m}$, defined by Eq.~\eqref{eq:defmvec}.
Let $t_{\vect{m}}$ denote a translation operator for the lattice vector
$\vect{m}$, defined by an application to a function $f$:
\begin{equation}
	(t_{\vect{m}}f)(\vect{r}) \equiv f(\vect{r}-\vect{m}).
\label{eq:translOperDef}
\end{equation}
An operator $A$ is \emph{translationally invariant} iff it commutes with the
translation operators for all lattice vectors $\vect{m}$ ($[\cdot,\cdot]$
denotes a commutator):
\begin{equation}
	\left[ A, t_{\vect{m}}\right] = 0.
\label{eq:translInvarOperCommut}
\end{equation}
Clearly, the momentum operator $\vect{p}$, as well as the spin operator
$\vect{\sigma}$ are translationally invariant. As a consequence, the composite
operators $p^2/2$ (nonrelativistic kinetic energy) and
$\vect{\sigma}\cdot\vect{p}$ are also translationally invariant. For this
reason, we can omit the spin- and momentum-dependence of an operator $A$ from
the following discussion without loss of generality. Let $A(\vect{r})$ be the
coordinate representation of $A$. Translation invariance of $A$
[Eq.~\eqref{eq:translInvarOperCommut}] then requires
\begin{equation}
	A(\vect{r}+\vect{m}) = A(\vect{r}).
\label{eq:translInvarOperCoord}
\end{equation}
Matrix elements of $A$ expressed in the discrete real-space basis of
Eq.~\eqref{eq:AObasisPeriodic} are obtained as
\begin{equation}
	A_{\mu\vect{m},\mu'\vect{m}'} = \int_{\reals^3} \dg{\chi}_{\mu\vect{m}}(\vect{r}) A(\vect{r})\chi_{\mu'\vect{m}'}(\vect{r}) \dV{r}.
\label{eq:rSpaceOp}
\end{equation}
For any lattice vector $\vect{n}$, it follows, that
\begin{equation}
	A_{\mu\vect{m},\mu'\vect{m}'} = A_{\mu\vect{m}+\vect{n},\mu'\vect{m}'+\vect{n}} = A_{\mu\vect{0},\mu'\vect{m}'-\vect{m}},
\label{eq:translInvarOperrSpace}
\end{equation}
implying that the real-space matrix elements of translationally invariant
operators have a Toeplitz structure. In addition, if the operator $A$ is
Hermitian, then
\begin{equation}
	\dg{A}_{\mu\vect{0},\mu'\vect{m}} = A_{\mu'\vect{0},\mu-\vect{m}},
\label{eq:translInvarOperrSpaceHermit}
\end{equation}
where $\dg{A}$ denotes the Hermitian conjugate within the $4\times 4$ bispinor
space.

Reciprocal-space elements of $A$ for $\vect{k},\vect{k}'\in\mathcal{K}$ are
acquired by using Eq.~\eqref{eq:kSpaceBasis} together with
Eq.~\eqref{eq:translInvarOperrSpace}:
\begin{equation*}
	A_{\mu\mu'}(\vect{k},\vect{k}') = \frac{1}{|\mathcal{K}|} \sum_{\vect{mm}'} \fourierM{k}{m}e^{i\vect{k}'\cdot\vect{m}'} A_{\mu\vect{0},\mu'\vect{m}'-\vect{m}}.
\end{equation*}
Changing the summation variables yields
\begin{align}
	A_{\mu\mu'}(\vect{k},\vect{k}') &= \delta(\vect{k}-\vect{k}') A_{\mu\mu'}(\vect{k}), \label{eq:translInvarOperkSpace1}\\
	A_{\mu\mu'}(\vect{k}) &= \sum_{\vect{m}} \fourierP{k}{m} A_{\mu\vect{0},\mu'\vect{m}}, \label{eq:translInvarOperkSpace2}
\end{align}
where we have employed
\begin{equation}
	\delta(\vect{k}) \equiv \frac{1}{|\mathcal{K}|} \sum_{\vect{m}} \fourierP{k}{m},
\label{eq:deltaFourierKernel}
\end{equation}
which is the \emph{Fourier kernel} representation of the Dirac delta
function. Notice that the symmetry in
Eq.~\eqref{eq:translInvarOperrSpace} resulted in the block-diagonal
reciprocal-space matrix [Eq.~\eqref{eq:translInvarOperkSpace1}]. This argument
can also be reversed, {\it i.e.\/} any block-diagonal $\vect{k}$-space matrix
will have a Toeplitz structure [Eq.~\eqref{eq:translInvarOperrSpace}] in real
space. We have applied this argument when constructing only the nonequivalent
elements of the real-space density matrix in
Eq.~\eqref{eq:rSpaceDmatQuadrature}. Finally, the symmetry in
Eq.~\eqref{eq:translInvarOperrSpaceHermit} leads to matrices in the reciprocal
space that are Hermitian for each $\vect{k}$ individually:
\begin{equation}
	\dg{A}_{\mu\mu'}(\vect{k}) = A_{\mu'\mu}(\vect{k}).
\label{eq:translInvarOperkSpaceHermit}
\end{equation}
Therefore, provided that the Fock matrix in Eq.~\eqref{eq:rSpaceFock4cMat}
satisfies the combined translational and Hermitian symmetry in
Eq.~\eqref{eq:translInvarOperrSpaceHermit}, the eigenvalues
$\epsilon(\vect{k})$ in Eq.~\eqref{eq:DKSinMatrixForm} are guaranteed to be
real. 

Translational symmetry allows us to assign finite expectation values of
operators that naturally describe extensive properties, such as the kinetic
energy of electrons. Beginning with a divergent expression for the expectation
value of a translationally invariant one-electron operator $A$ (given that the
density matrix is translationally invariant as well), we can write
\begin{align*}
	\mean{A} &= \sum_{\vect{mm}'} \Tr\left[ A_{\mu\vect{m},\mu'\vect{m}'}D^{\mu'\vect{m}',\mu\vect{m}} \right] \\
	&= \sum_{\vect{m}}1 \sum_{\vect{m}'} \Tr\left[ A_{\mu\vect{0},\mu'\vect{m}'}D^{\mu'\vect{m}',\mu\vect{0}} \right],
\end{align*}
where $\Tr$ denotes the trace in the $4\times 4$ bispinor space. If we employ
the short-hand notation from Eq.~\eqref{eq:rSpaceTrace}, and realize, that
$\sum_{\vect{m}}1\equiv N$ is the total (infinite) number of unit cells, we can
calculate the expectation value of $A$ \emph{per unit cell} in the
thermodynamic limit ($N\rightarrow\infty$) as
\begin{equation}
	\frac{\mean{A}}{N} = \Tr\left[ A_uD^{\bar{u}} \right].
\label{eq:meanOfTranslInvarOp}
\end{equation}

\section{\label{app:SphericalMultipoleExpansion}Spherical multipole expansion}

Here, we summarize the formulation of the spherical multipole expansion needed
to evaluate the far-field contribution to the Coulomb operator. We follow the
framework of Helgaker \emph{et al.}~\cite{Helgaker-405-2000} and Watson
\emph{et al.}~\cite{Watson-JCP-121-2915-2004} The Coulomb interaction operator
$|\vect{r}_1-\vect{r}_2|^{-1}\equiv r_{12}^{-1}$ can be expanded (as a function
of 6 variables) around an arbitrary center $(\vect{P},\vect{Q})$ into a
spherical multipole expansion which takes the form
\begin{widetext}
\begin{equation}
	\frac{1}{r_{12}} = \sum_{l=0}^{\infty}\sum_{m=-l}^{l}\sum_{j=0}^{\infty}\sum_{k=-j}^{j} R^{lm}(\vect{r}_1-\vect{P})\Theta_{lm,jk}(\vect{Q}-\vect{P})R^{jk}(\vect{r}_2-\vect{Q}),
\label{eq:SMEFirst}
\end{equation}
\end{widetext}
where
\begin{equation}
	\Theta_{lm,jk}(\vect{R}) = (-1)^j I^{*}_{l+j,m+k}(\vect{R}),
\label{eq:SMEInteractionTensor}
\end{equation}
is the \emph{interaction tensor}, $R^{lm}(\vect{r})$ and $I_{lm}(\vect{r})$ are
the \emph{scaled regular} and \emph{scaled irregular solid harmonics},
respectively, defined as
\begin{align}
	R^{lm}(\vect{r}) &= \frac{1}{\sqrt{(l-m)!(l+m)!}} r^l C_{lm}(\theta,\varphi), \label{eq:SMEsolidHarmR}\\
	I_{lm}(\vect{r}) &= \sqrt{(l-m)!(l+m)!} r^{-l-1} C_{lm}(\theta,\varphi). \label{eq:SMEsolidHarmI}
\end{align}
Here $C_{lm}(\theta,\varphi)$ are eigenfunctions of the angular momentum
operators $L^2,L_z$, namely, the \emph{spherical harmonics} in Racah's
normalization, obtained from the conventional spherical harmonics
$Y_{lm}(\theta,\varphi)$ as
\begin{equation}
	C_{lm}(\theta,\varphi) = \sqrt{\frac{4\pi}{2l+1}} Y_{lm}(\theta,\varphi).
\label{eq:SMESphericalHarmRacah}
\end{equation}
We shall use the compact matrix notation
\begin{equation}
	\frac{1}{r_{12}} = R^T(\vect{r}_1-\vect{P})\Theta(\vect{Q}-\vect{P})R(\vect{r}_2-\vect{Q}),
\label{eq:SMECompact}
\end{equation}
where $R$ is a vector and $\Theta$ is a matrix defined by their respective
elements $R^{lm}$ and $\Theta_{lm,jk}$.  The series in
Eq.~\eqref{eq:SMECompact} is convergent for all points
$(\vect{r}_1,\vect{r}_2)$ that satisfy the condition
\begin{equation}
	|\vect{r}_1 - \vect{r}_2 + \vect{Q} - \vect{P}| < |\vect{Q} - \vect{P}|.
\label{eq:SMEConvCondition}
\end{equation}
The scaled regular and irregular solid harmonics have the following properties
($\lambda\in\reals$ is an arbitrary scaling factor):
\begin{subequations}
\label{eq:SMESolidHarmPropertiesGroup}
\begin{align}
	R^{l-m}(\vect{r}) &= (-1)^m R^{lm*}(\vect{r}),  \label{eq:SMESolidHarmPropertiesR1}\\
	I_{l-m}(\vect{r}) &= (-1)^m I_{lm}^*(\vect{r}), \label{eq:SMESolidHarmPropertiesI1}\\
	R^{lm}(\lambda\vect{r}) &= \lambda^l R^{lm}(\vect{r}), \label{eq:SMESolidHarmPropertiesR2}\\
	I_{lm}(\lambda\vect{r}) &= \frac{1}{|\lambda|}\frac{1}{\lambda^l} I_{lm}(\vect{r}). \label{eq:SMESolidHarmPropertiesI2}
\end{align}
\end{subequations}
The regular solid harmonics obey the addition theorem
\begin{equation}
	R^{lm}(\vect{r}-\vect{P}) = \sum_{j=0}^l\sum_{k=-j}^j R_{l-j,m-k}(-\vect{P}) R^{jk}(\vect{r}),
\label{eq:SMEregHarmTransl}
\end{equation}
which can be written in the following matrix form
\begin{equation}
	R(\vect{r}-\vect{P}) = W(\vect{P})R(\vect{r}),
\label{eq:SMEregHarmTranslCompact}
\end{equation}
where $W$ is the \emph{translation tensor}, its elements defined as
\begin{equation}
	W_{lm,jk}(\vect{P}) = R_{l-j,m-k}(-\vect{P}).
\label{eq:SMEtranslTensor}
\end{equation}
The translation tensor $W$ can be used to evaluate the regular solid harmonics
for shifted arguments. Moreover, we can apply
Eq.~\eqref{eq:SMEregHarmTranslCompact} to derive a similar rule for the
interaction tensor. Multipole expansions of $r_{12}^{-1}$ expanded around 2
different centers $(\vect{P},\vect{Q})$ and $(\bar{\vect{P}},\bar{\vect{Q})}$
must coincide, so that
\begin{align*}
	\frac{1}{r_{12}} &= R^T(\vect{r}_1-\vect{P})\Theta(\vect{Q}-\vect{P})R(\vect{r}_2-\vect{Q}) \\
	&= R^T(\vect{r}_1-\bar{\vect{P}})\Theta(\bar{\vect{Q}}-\bar{\vect{P}})R(\vect{r}_2-\bar{\vect{Q}}).
\end{align*}
Applying the addition theorem [Eq.~\eqref{eq:SMEregHarmTranslCompact}], we
identify
\begin{equation}
	\Theta(\vect{Q}-\vect{P}) = W^T(\bar{\vect{P}}-\vect{P})\Theta(\bar{\vect{Q}}-\bar{\vect{P}})W(\bar{\vect{Q}}-\vect{Q}).
\label{eq:SMEInteractionTensorTransl1}
\end{equation}
Using $W(\vect{0})=\idmatrix$, and setting $\bar{\vect{P}}=\vect{P}$ and
$\bar{\vect{Q}}=\vect{P}+\vect{Q}$ in
Eq.~\eqref{eq:SMEInteractionTensorTransl1}, we obtain the corollary
\begin{equation}
	\Theta(\vect{Q}-\vect{P}) = \Theta(\vect{Q})W(\vect{P}).
\label{eq:SMEInteractionTensorTransl2}
\end{equation}
In the present implementation, we avoid using complex numbers for multipole
expansions by expressing interaction and translation tensors in terms of the
real (regular and irregular) solid harmonics, which we construct from
recurrence equations (see Ref.~\cite{Helgaker-405-2000}) and we do
therefore not evaluate the zero imaginary part of the real-valued Coulomb
$r_{12}^{-1}$ operator.

\section{\label{app:LatticeSumOfInteractionTensors}Lattice sum of interaction tensors}

Here we prove the recurrence relation in Eq.~\eqref{eq:latSumRenormEq},
establishing a rapidly convergent scheme for the computation of  lattice sums
of spherical interaction tensors. Let us begin by fragmenting the far-field
(FF) into layers FF$_r$ as follows: Let the near-field (NF) be a block
consisting of unit cells with indices $n^i=-N_i,\ldots,N_i$ for each of the
periodic dimensions $i=1,\ldots,d$. For generic noncubic lattices, such an
object has a diamondlike shape. The first layer of the far-field, FF$_1$,
envelopes the NF by placing supercells in all directions, each supercell having
as many unit cells as the NF itself. The process is then repeated for the next
layer of the far-field, FF$_2$, with the exception that  the supercell now
contains all unit cells in both NF and FF$_1$, as  depicted in the following
scheme:
\begin{widetext}
\begin{equation}
	\ldots \mid \underbrace{\overbrace{-3N_i-1\ldots -N_i-1}^{\text{FF}_1}}_{2N_i+1} \mid \underbrace{\overbrace{-N_i\ldots -1}^\text{NF}\boxed{0}\overbrace{1\ldots N_i}^\text{NF}}_{2N_i+1} \mid \underbrace{\overbrace{N_i+1\ldots 3N_i+1}^{\text{FF}_1}}_{2N_i+1} \mid \underbrace{\overbrace{3N_i+2\ldots 9N_i+4}^{\text{FF}_2}}_{2(3N_i+1)+1} \mid \ldots
\label{eq:FFLayersScheme}
\end{equation}
\end{widetext}
Let $N_{ir}$ denote the upper extent of the far-field layer $r$ in the
direction $i$, \emph{i.e.} it is the index of the unit cell that is the
farthest from the center $0$. Then $N_{ir}$ satisfies the following recurrence
relations ($r=0$ labels the NF)
\begin{equation}
\begin{split}
	N_{i0} &= N_i, \\
	N_{ir+1} &= 3N_{ir} + 1,
\end{split}
\label{eq:FFLayersExtents}
\end{equation}
which have the solution
\begin{equation}
	N_{ir} = \frac{(2N_i+1)3^r - 1}{2}.
\label{eq:FFLayersExtentsSolution}
\end{equation}
The number of unit cells in layer $r$ is given by
\begin{equation}
	|\text{FF}_r| = 3^{d(r-1)}(3^d - 1) |\text{NF}|,
\label{eq:FFLayersNrElms}
\end{equation}
where $|X|$ denotes the number of elements of $X$. From
Eq.~\eqref{eq:FFLayersNrElms}, we can see that the sizes of the layers form a
geometric sequence. Therefore, the partitioning in
Eq.~\eqref{eq:FFLayersScheme} divides the space into regions that become
exponentially larger with each new layer.  Formally, we define FF$_r$ as
\begin{equation}
	\text{FF}_r = \left\{ (n^1\ldots n^d)\in\mathbb{Z}^d; 1\leq \max_{i=1\ldots d}\left(\frac{|n^i|-1}{N_{ir-1}}\right) \leq 3 \right\}.
\label{eq:FFlayers}
\end{equation}
The overall far-field is then given by the union
\begin{equation}
	\text{FF} = \bigcup_{r=1}^{\infty} \text{FF}_r,
\label{eq:FFWhole}
\end{equation}
and the lattice sum in Eq.~\eqref{eq:latSumOfIntTens} becomes
\begin{align*}
	\Lambda = \sum_{\vect{n}\in\text{FF}} \Theta(\vect{n}) = \lim_{t\rightarrow\infty} \sum_{r=1}^t \sum_{n\in\text{FF}_r} \Theta(n^i\vect{a}_i),
\end{align*}
where we have abbreviated the summation indices as $n=(n^1\ldots n^d)$.  It
follows, that the lattice sum is obtained as a limit of partial sums
\begin{align}
	\Lambda &= \lim_{t\rightarrow\infty} \Lambda^t, \label{eq:latSumAsLimitApp}\\
	\Lambda^t &= \sum_{r=1}^t \sum_{n\in\text{FF}_r} \Theta(n^i\vect{a}_i). \label{eq:latSumPartial}
\end{align}
Let us consider the term $t+1$:
\begin{align}
	\Lambda^{t+1} &= \Lambda^1 + \sum_{r=2}^{t+1} \sum_{n\in\text{FF}_r} \Theta(n^i\vect{a}_i) \notag\\
	&= \Lambda^1 + \sum_{r=1}^t \sum_{n\in\text{FF}_{r+1}} \Theta(n^i\vect{a}_i). \label{eq:latSumNextIter}
\end{align}
The following identity relates the two sums over different layers of the
far-field
\begin{equation}
	\sum_{n\in\text{FF}_{r+1}} \Theta(n^i\vect{a}_i) = \sum_{n\in\text{FF}_r} \sum_{\mu\in\mathcal{P}} \Theta\left((3n^i-\mu^i)\vect{a}_i\right),
\label{eq:latSumIdentityBetweenFFs}
\end{equation}
where $\mathcal{P}$ is the Cartesian power
\begin{equation*}
	\mathcal{P} = \left\{-1,0,1\right\}^d,
\end{equation*}
for $d=3, \mathcal{P}=\left\{(\pm 1,\pm 1,\pm 1),(\pm 1,\pm
1,0),\ldots\right\}$ and contains the reference unit cell and all its 26
nearest neighbours. 

Up to this point, the proof has been of a general nature --- we did not need to
specify $\Theta$ or use its properties. However, in order to obtain an
applicable recursive formulation, we need to express the term $\Lambda^{t+1}$
via the previous terms. To proceed, we therefore apply the addition theorem in
Eq.~\eqref{eq:SMEInteractionTensorTransl2},  factorizing the interaction tensor
as
\begin{align*}
	\Theta\left((3n^i-\mu^i)\vect{a}_i\right) &= \Theta\left(3n^i\vect{a}_i\right) W\left(\mu^i\vect{a}_i\right) \\
	&\equiv \mathcal{U}\left[\Theta\left(n^i\vect{a}_i\right)\right] W\left(\mu^i\vect{a}_i\right),
\end{align*}
where $W$ is the translation tensor [Eq.~\eqref{eq:SMEtranslTensor}], and where
we have defined the scaling operator $\mathcal{U}$ as
\begin{equation}
	\mathcal{U}\left[\Theta_{lm,jk}(\vect{n})\right] \equiv \Theta_{lm,jk}(3\vect{n}) = \frac{1}{3^{l+j+1}}\Theta_{lm,jk}(\vect{n}).
\label{eq:latSumScalingOpApp}
\end{equation}
Here we applied the scaling property of the irregular solid harmonics
[Eq.~\eqref{eq:SMESolidHarmPropertiesI2}]. Returning to
Eq.~\eqref{eq:latSumNextIter}, this leads to
\begin{align*}
	\Lambda^{t+1} &= \Lambda^1 + \sum_{r=1}^t \sum_{n\in\text{FF}_r} \sum_{\mu\in\mathcal{P}} \Theta\left((3n^i-\mu^i)\vect{a}_i\right) \\
	&= \Lambda^1 + \mathcal{U}\left[ \sum_{r=1}^t \sum_{n\in\text{FF}_r} \Theta\left(n^i\vect{a}_i\right) \right] \sum_{\mu\in\mathcal{P}} W\left(\mu^i\vect{a}_i\right).
\end{align*}
If we define the aggregate translation matrix
\begin{equation}
	\mathcal{W} = \sum_{\mu\in\mathcal{P}} W\left(\mu^i\vect{a}_i\right) \equiv \sum_{\mu^1\ldots \mu^d=-1}^1 W(\mu^i\vect{a}_i),
\label{eq:latSumTranslSumApp}
\end{equation}
then
\begin{equation}
	\Lambda^{t+1} = \Lambda^1 + \mathcal{U}\left(\Lambda^t\right)\mathcal{W},
\label{eq:latSumRenormEqApp}
\end{equation}
which completes the proof.


\bibliography{references}

\end{document}